\documentclass[prd,preprint,tightenlines,floatfix,showpacs,preprintnumbers,nofootinbib,eqsecnum]{revtex4}

 \usepackage[dvips,final]{graphicx}
  \usepackage{amssymb}
   \usepackage{amsmath}
    \usepackage{amsfonts}
     \usepackage{epsfig}
      \usepackage{bm}

\begin{document}

\begin{flushright}
LU TP 13-27\\
October 2013
\end{flushright}

\title{Vector-like technineutron Dark Matter:\\
is a QCD-type Technicolor ruled out by XENON100?}

\author{Roman Pasechnik}
\email{Roman.Pasechnik@thep.lu.se} \affiliation{Department of
Astronomy and Theoretical Physics, Lund University, SE-223 62 Lund,
Sweden}

\author{Vitaly Beylin}
\affiliation{Research Institute of Physics, Southern Federal
University, 344090 Rostov-on-Don, Russian Federation}

\author{Vladimir Kuksa}
\affiliation{Research Institute of Physics, Southern Federal
University, 344090 Rostov-on-Don, Russian Federation}

\author{Grigory Vereshkov}
\affiliation{Research Institute of Physics, Southern Federal
University, 344090 Rostov-on-Don, Russian Federation}
\affiliation{Institute for Nuclear Research of Russian Academy of
Sciences, 117312 Moscow, Russian Federation\vspace{1cm}}

\begin{abstract}
\vspace{0.5cm} We continue to explore a question about the existence
of a new strongly coupled dynamics above the electroweak scale. The
latter has been recently realized in the simplest consistent
scenario, the vector-like (or chiral-symmetric) Technicolor model
based upon the gauged linear $\sigma$-model. One of the predictions
of a new strong dynamics in this model, the existence of stable
vector-like technibaryon states at a TeV scale, such that the
lightest neutral one could serve as a Dark Matter candidate. Here,
we consider the QCD-type Technicolor with $SU(3)_{\rm TC}$ confined
group and one $SU(2)_{\rm W}$ doublet of vector-like techniquarks
and test this model against existing Dark Matter astrophysics data.
We show that the spin-independent Dirac technineutron-nucleon cross
section is by far too large and ruled out by XENON100 data. We
conclude that vector-like techniquark sectors with an odd group of
confinement $SU(2n+1)_{\rm TC},$ $n=1,2,\dots$ and with ordinary
vector-like weak $SU(2)_{\rm W}$ interactions are excluded if the
technibaryon number is conserved. We discuss a possible generic TC
scenario with technibaryon sector interacting via an extra vector
$SU(2)_{\rm V}$ other than the standard weak $SU(2)_{\rm W}$ and
consider immediate implications for the cosmological evolution and
freeze out of heavy relic technineutrons.
\end{abstract}


\maketitle

\section{Introduction}

The undoubtful existence of the Dark Matter (DM) comprising about a
third (or more precisely, about $27$ \% \cite{Planck}) of total mass
of the Universe today remains the strongest phenomenological
evidence for New Physics beyond the Standard Model (SM) required by
astrophysics measurements. The hypothetical weakly-interacting
massive particles (WIMPs), the DM is possibly composed of, and their
properties are yet undiscovered at the fundamental level while the
DM itself is being regarded as one of the major cornerstones of
modern theoretical astrophysics and cosmology \cite{Trimble:1987ee}.
Such an uneasy situation motivates ongoing search for appropriate
Particle Physics candidates for WIMPs away from constantly improving
observational bounds.

Traditionally, lightest supersymmetric particles (LSPs) predicted by
supersymmetry (SUSY) \cite{SUSY} such as neutralino are often
referred to as to the best DM candidates \cite{Jungman:1995df}, and
this is considered to be one of the major advantages of SUSY-based
SM extensions (for an overview of existing DM candidates, see e.g.
Refs.~\cite{Bertone:2004pz,Roszkowski:2004jc} and references
therein). Direct SUSY searches are currently ongoing at the LHC and
major direct and indirect DM detection experiments, so that the
parameter space of simplest SUSY scenarios is getting more and more
constrained (for the most recent exclusion limits and their effects
on SUSY DM candidates see e.g. Ref.~\cite{SUSY-DM-constraints}).

In this paper, we consider one of the alternatives to SUSY-based DM
candidates predicted by dynamical electroweak symmetry breaking
(EWSB) and compositeness scenarios, the lightest heavy neutral
technibaryon (or T-baryon) state $N$. In case of the odd QCD-type
$SU(3)_{\rm TC}$ group of confinement extending the SM gauge group
such a candidate is often referred to as the Dirac T-neutron in
analogy to ordinary neutrons from low energy hadron physics. The
idea of composite DM candidates has a long history starting from
mid-eighties from Ref.~{\cite{Nussinov}} where it has been claimed
that an excess of T-baryons possibly built up in the early Universe
can explain the observed missing mass. So far, a number of different
models of composite DM candidates and hypotheses about their origin
and interactions has been proposed. Generic DM signatures from
Technicolor-based models with stable T-baryons were discussed e.g.
in Refs.~\cite{Gudnason,Khlopov,DelNobile:2011uf} (for a review see
also Ref.~\cite{Sannino:2009za} and references therein). In
particular, well-known minimal dynamical EWSB mechanisms predict
relatively light T-baryon states as pseudo Nambu-Goldstone bosons of
the underlying gauge theory \cite{Belyaev:2010kp,Ryttov}. The latter
can naturally provide asymmetric DM candidates if one assumes the
existence of a T-baryon asymmetry in Nature similarly to ordinary
baryon asymmetry \cite{Petraki:2013wwa}. Having similar mechanisms
for ordinary matter and DM formation in early Universe one would
expect the DM density to be of the same order of magnitude as that
of baryons. Depending on a particular realization of dynamical EWSB
mechanism such composite DM candidates may be self-interacting which
helps in avoiding problematic cusp-like DM halo profiles
\cite{Kouvaris:2013gya}.

All of the existing composite DM models rely on the basic assumption
about New Physics extension of the SM by means of extra confined
matter sectors. These ideas were realized in a multitude of
Technicolor (TC) models developed so far \cite{TC} (for a detailed
review on the existing TC models, see e.g.
Refs.~\cite{Hill:2002ap,Sannino}). Historically, the first TC models
with dynamical EWSB are based upon the idea that the Goldstone
degrees of freedom (technipions or T-pions) appearing after the
global chiral symmetry breaking $SU(2)_{\rm L}\otimes SU(2)_{\rm
R}\to SU(2)_{\rm V}$ are absorbed by the SM weak gauge bosons which
thereby gain masses. The dynamical EWSB mechanism is then triggered
by the condensate of fundamental technifermions (or T-quarks) in
confinement, $\langle {\tilde Q}\bar{\tilde Q}\rangle\not=0$.
Traditional TC models with dynamical EWSB are faced with the problem
of the mass generation of standard fermions, which was consistently
resolved in the Extended TC model \cite{Extended-TC}. However, many
of the existing TC-based models have got severely constrained or
often ruled out by the EW precision data \cite{Peskin:1990zt}.
Generally, in these schemes noticeable contributions to strongly
constrained Flavor Changing Neutral Current (FCNC) processes appear
together with too large contributions to Peskin-Takeuchi
(especially, to $S$) parameters. Further developments of the TC
ideas have resulted in the Walking TC and the vector-like (or
chiral-symmetric) TC which succeeded in resolving the
above-mentioned problems and remain viable scenarios of the
dynamical EWSB
\cite{Appelquist:1986an,Sannino:2004qp,Foadi:2007ue,our-TC}.

In this paper, we continue investigation of promising
phenomenological implications of the vector-like TC model proposed
recently in Ref.~\cite{our-TC}. This is one of the simplest
successful realizations of the bosonic TC scenarios -- an extension
of the SM above the electroweak (EW) scale which includes both a
Higgs doublet $H$ and a new strongly-coupled {\it vector-like}
techniquark sector (for different realizations of the bosonic TC
ideas, see e.g.
Refs.~\cite{Simmons:1988fu,Samuel:1990dq,Kagan:1991gh,bos-TC}). In
contrast to conventional (Extended and Walking) TC models, in the
vector-like TC model the mechanism of the EWSB and generation of SM
fermions masses is driven by the Higgs vacuum expectation value
(vev) in the standard way, irrespectively of (elementary or
composite) nature of the Higgs field itself. Similarly to other
bosonic TC models, the Higgs field $H$ develops a vev which in turn
is induced by the T-quark condensate. Thus, it is possible to
assimilate the SM-like Higgs boson while the Higgs vev acquires a
natural interpretation in terms of the T-quark condensates. This
means the Higgs mechanism is not the primary source of the EWSB, but
effectively induced by an unknown TC dynamics at high scales.

The vector-like TC model \cite{our-TC} is based upon
phenomenologically successful gauged linear $\sigma$-model
(GL$\sigma$M) initially proposed in Ref.~\cite{Lee} and further
elaborated in Refs.~\cite{LSigM,SU2LR}. It is well-known that in the
low energy limit of QCD and in the limit of massless $u$ and $d$
quarks, the resulting QCD Lagrangian with switched off weak
interactions of $u,d$ quarks possesses {\it exact global chiral}
$SU(2)_{\rm L}\otimes SU(2)_{\rm R}$ symmetry. The physical degrees
of freedom in this Lagrangian are given by a superposition of
initially chiral quark fields -- {\it the Dirac $u,d$-quark fields}.
Global $SU(2)_{\rm L}\otimes SU(2)_{\rm R}$ is then considered as
{\it a classification symmetry of composite states} giving rise to
the lightest hadrons in the physical spectrum and nicely predicting
their properties. This model predicts the lightest {\it physical}
pseudoscalar T-pion $\tilde{\pi}$, scalar T-sigma $\tilde{\sigma}$
fields as well as T-baryon states classified according to
representations of gauged vector subgroup $SU(2)_{\rm V\equiv L+R}$
of original global chiral $SU(2)_{\rm L}\otimes SU(2)_{\rm R}$
symmetry. Its complete gauging is also possible at the composite
level giving rise to effective field theory describing the
``chiral-gauge'' interactions between bound states in adjoint (e.g.
composite vector/pseudovector fields and pions) and fundamental
(e.g. composite baryons, constituent ``dressed-up'' quarks)
representations. But this gauging makes sense {\it only} at the
level of bound states, but never at the fundamental level.

As usual, we consider the spontaneous chiral symmetry breaking in
the T-hadron sector happens in the chiral-symmetric (vector-like)
way
\begin{eqnarray}
SU(2)_{\rm L}\otimes SU(2)_{\rm R}\to SU(2)_{\rm V\equiv L+R}\,.
\label{CSB}
\end{eqnarray}
In Ref.~\cite{our-TC} it was argued that in the low energy limit the
vector-like gauge group $SU(2)_{\rm V}$ should be {\it identified}
with the weak isospin group $SU(2)_{\rm W}$ of the SM,
i.e.\footnote{In addition, in this paper we explore a possible
option when gauged $SU(2)_{\rm V}$ is associated with another
fundamental gauge group in the T-quark sector different from the SM
weak isospin group such that corresponding gauge bosons $Z',W'$ are
very heavy and their mixing with SM gauge bosons is strongly
suppressed. For example, this group can be the one coming from the
LR-symmetric generalization of the SM at high scales.}
\begin{eqnarray}
SU(2)_{\rm V\equiv L+R}\simeq SU(2)_{\rm W}\,. \label{ident}
\end{eqnarray}
Such a ``gauging'' of the vector subgroup $SU(2)_{\rm V\equiv L+R}$
and its identification with the SM gauge isospin group do not mean
that one introduces extra elementary gauge bosons to the existing
fundamental theory, e.g. to the SM or its possible high-scale gauge
extensions. In our context, procedure (\ref{ident}) means the very
simple thing: both T-quarks and T-hadrons interact with {\it already
existing} gauge bosons in the SM in the low-energy effective field
theory limit with local gauge couplings \cite{our-TC}. Note, such an
identification is automatic at the fundamental T-quark level -- the
Dirac T-quarks reside in the weak isospin group $SU(2)_{\rm W}$ from
the beginning. Indeed, in the high energy limit of the theory, the
global classification symmetry $SU(2)_{\rm L}\otimes SU(2)_{\rm R}$
is restored, while Dirac T-quark fields, along with chiral SM
fermion fields, reside in fundamental representations of the SM
gauge $SU(2)_{\rm W}$ and no extra fundamental gauge bosons are
needed.

The most critical part of the proposal is that the resulting
unbroken local chiral-symmetric subgroup $SU(2)_{\rm V\equiv L+R}$
describes the gauge interactions of constituent T-quarks and
T-hadrons with local gauge couplings {\it in the low energy limit of
the effective field theory} especially interesting for
phenomenology. As of the primary goal of this work, we would like to
test against astrophysics DM data if these interactions are ordinary
weak or not under an assumption for an odd $SU(3)_{\rm TC}$ group in
confinement.

One should remember that identification of the local vector subgroup
of the chiral group with the SM weak isospin group (\ref{ident}) is
a purely phenomenological procedure which leads to correct results
in the low energy limit of the theory. In reality, of course, the
global classification T-flavor group $SU_{\rm L}(2)\otimes SU_{\rm
R}(2)$ has nothing to do with the EW gauge group of the SM. At the
first stage, the T-flavor group is used for classification of {\it
composite} T-hadrons and, in particular, predicts the existence of
T-pions, T-sigma and T-baryons states. At the second stage, one
notices that T-quarks entering the composite T-hadrons besides
T-strong interactions participate also in the fundamental EW
interactions. One should therefore calculate the {\it EW form
factors} of composite T-hadrons. The corresponding EW interactions
must then be also introduced at {\it the fundamental T-quark level}
consistently with those at {\it the composite level T-hadron level}.
At the third stage, in the phenomenologically interesting low-energy
limit of the theory the EW form factors approach the renormalized EW
constants (since the T-hadron substructure does not emerge at
relatively small momentum transfers). The latter should be
calculated after {\it reclassification} of T-hadrons under the EW
group representations. This three-fold generic scheme will be used
below for description of EW interactions of T-hadrons.

As one of the important features of the VLTC model, after the chiral
symmetry breaking in the T-quark sector the left and right
components of the original Dirac T-quark fields can interact with
the SM weak $SU(2)_{\rm W}$ gauge bosons with vector-like couplings,
in opposition to ordinary SM fermions, which interact under
$SU(2)_{\rm W}$ by means of their left-handed components only.

Remarkably enough, in this model the oblique (Peskin-Takeuchi)
parameters and FCNC corrections turn out to be naturally very small
and fully consistent with the current EW constraints as well as with
the most recent Higgs couplings measurements at the LHC in the limit
of small Higgs-T-sigma mixing. Most importantly, this happens
naturally in the standard quantum-field theory framework implemented
in rigorous quark-meson approaches of hadron physics without
attracting any extra holographic or other special arguments from
unknown high-scale physics. For simplicity, here we adopt the
simplest version of the Standard Model with one Higgs doublet, and
the question whether it is elementary or composite is not critical
for further considerations. The new heavy physical states of the
model (additional to those in the SM) are the singlet T-sigma
$\tilde{\sigma}$, triplet of T-pions $\tilde{\pi}_a,\,a=1,2,3$, and
constituent T-quarks $\tilde{Q}$ which acquire masses via the
T-quark condensate as an external source and the T-sigma vev. We are
focused on phenomenological studies of such a low energy effective
field theory at typical momentum transfers squared $Q^2\ll
\Lambda_{\rm TC}^2$ in the considered linear $\sigma$-model
framework without attempting to construct a high energy unifying new
strongly coupled dynamics with the SM at the moment.

Note also that the vector-like TC model offers a simple method of
phenomenologically consistent construction of the vector-like
ultraviolet completion of a strongly-coupled theory which can be
further exploited in the composite Higgs models as well as in
attempts for Grand-like TC unification with the SM at high scales
(see e.g. Refs.~\cite{Marzocca:2012zn,Caracciolo:2012je} and
references therein).

The VLTC scenario represents the very first step focussing on the
low-energy implications of a new strongly coupled dynamics with
chiral-symmetric UV completion -- the first relevant step for
searches for such a dynamics at the LHC and in astrophysics -- {\it
formally} keeping the elementary Higgs boson as it is in the
one-doublet SM which does not satisfy the naturalness criterium.
From the theoretical point of view, the model points out a promising
path towards a consistent formulation of composite Higgs models in
extended chiral-gauge theories with chiral-symmetric UV completion.
Most importantly, such a strongly coupled sector survives the EW
precision tests with minimal vector-like confined sector ($U$ and
$D$ T-quarks) without any extra assumptions. Excluding the
naturalness criterium, other three important points which are
considered to be primary achievements of the VLTC model
\cite{our-TC} can be summarized as follows:
\begin{itemize}
\item The effective Higgs mechanism of dynamical EW symmetry
breaking in the conformal limit of the theory forbidding Higgs
$\mu$-terms is naturally emerged in this approach. The Higgs vev is
automatically expressed in terms of the T-quark condensate such that
the EW symmetry is broken simultaneously with the chiral symmetry
breaking. No T-pions are eaten and remain physical, they escape
current detection limits due to extremely suppressed loop-induced
couplings to two or even three gauge bosons (depending on T-quark
hypercharge and TC gauge group) at leading order only, and can
remain very light.
\item The phenomenologically and theoretically consistent
minimal vector-like UV completion based upon the linear
$\sigma$-model with the global chiral $SU_{\rm L}(2)\otimes SU_{\rm
R}(2)$ symmetry group is proposed. In the minimal VLTC, the model
works perfectly with only two vector-like T-flavors easily passing
the EW constraints and (almost) standard Higgs couplings without any
extra assumptions. The model is capable of unique predictions of
possibly small Higgs couplings deviations from the standard ones.
\item There are specific phenomenological consequences of
such a new dynamics at the LHC, e.g. light hardly detectable
technipions with multi-boson final states produced via a suppressed
VBF only, and possible distortions of the Higgs boson couplings and
especially self-couplings, vector-like T-baryon states at the LHC
with a large missing-$E_T$ and asymmetry signatures as well as
implications for Cosmology (vector-like T-baryon Dark Matter).
\end{itemize}

The proposed scenario, at least, in its simplest form discussed in
Ref.~\cite{our-TC}, does not attempt to resolve the
naturalness/hierarchy problem of the SM and does not offer a
mechanism for generation of current T-quark masses. It is considered
as a low-energy phenomenologically consistent limit of a more
general strongly-coupled dynamics which is yet to be constructed (it
has the same status as the low-energy effective field theories
existing in hadron physics).

Making three above points work coherently together in a single
phenomenologically motivated model is the first important step
towards a consistent high-energy description of vector-like
Technicolor dynamics made in Ref.~\cite{our-TC}. And here we aim at
analysis of immediate implications of this scenario in Dark Matter
searches. New heavy vector-like T-baryon states, T-proton $P$ and
T-neutron $N$ states, at a TeV mass scale are naturally predicted
and introduced in a similar way as in low-energy hadron physics
allowing for a possible interpretation of the DM in the considering
framework. A thorough analysis of distinct features of the T-neutron
DM with generic weak-type $SU(2)_{\rm V}$ interactions due to a
vector-like character of T-neutron gauge interactions and additional
T-strong channels (via T-pion/T-sigma) along with existing direct DM
detection constraints is the primary goal of our current study.

Even though the EW precision constraints are satisfied for any
$SU(n)_{\rm TC}$ group with vector-like weak interactions
\cite{our-TC}, it is still an open question, if astrophysics
constraints are satisfied for any $SU(n)_{\rm TC}$ group as well.
The DM exclusion limits, therefore, become an extra important source
of information about TC dynamics which has a power to constrain the
parameter space of the vector-like TC model even more. One of the
unknowns we would like to consider here is the rank of the confined
group. In particular, we will discuss for which $SU(n)_{\rm TC}$
groups in confinement it is possible to make the identification of
the gauge groups (\ref{ident}) in the T-quark/T-baryon sectors, and
for which -- it is not, based upon existing constraints from DM
astrophysics. The latter will be our main conclusion of this work.

The paper is organized as follows. Section II provides a brief
overview of vector-like $SU(3)_{\rm TC}$ TC model with Dirac
T-baryons with generic weak-type $SU(2)$ interactions (before
identification (\ref{ident})). Section III contains a discussion of
the T-baryon mass spectrum, in particular, important mass splitting
between T-proton and T-neutron. In Section IV, we consider typical
T-baryon annihilation processes in the cosmological plasma in two
different cases -- in the high- and low-symmetry phases. Section V
is devoted to a discussion of cosmological evolution of T-neutrons
in two cases of symmetric and asymmetric DM. In Section VI, major
implications of direct detection limits to the considering
vector-like T-neutron DM model are outlined. It was shown that
weakly $SU(2)_{\rm W}$ interacting Dirac vector-like T-baryons are
excluded by recent XENON100 data \cite{Aprile:2012nq}, which poses
an important constraint on the rank of the confined group in the
T-quark sector under condition (\ref{ident}). Finally, Section VII
contains basic concluding remarks.

\section{Vector-like T-baryon interactions}

As one of the basic predictions of the vector-like and other bosonic
TC models, the EWSB occurs by means of the effective Higgs mechanism
induced by a condensation of confined fermions. The basic hypothesis
which should be thoroughly tested against both astrophysical
(primarily, DM) and collider (new exotic lightest T-hadron states)
data can be formulated as follows: {\it the energy scales of the
EWSB and T-confinement have a common quantum-topological nature and
are determined by a non-perturbative dynamics of the
T-quark--T-gluon condensate}. This work aims at testing this
hypothesis against DM astrophysics data.

The key difference of the vector-like TC approach earlier developed
in Ref.~\cite{our-TC} from other known dynamical EWSB mechanisms is
primarily in the vector-like character of gauge interactions of
T-quarks as a natural consequence of the local chiral symmetry
breaking (\ref{CSB}) and a possible identification of the local
chiral vector-like subgroup with the weak isospin group of the SM
(\ref{ident}). The latter requirement strongly reduces extra loop
contributions to the EW observables and the new dynamics may slip
away from the EW precision tests and ongoing Higgs couplings studies
at the LHC even in the case of relative proximity of the
T-confinement $\Lambda_{\rm TC}$ and EW $M_{\rm EW}\sim 100$ GeV
scales. So, frequent references to ``Technicolor'' as to a ``dead
concept'' in the literature do not apply to the vector-like TC
model, at least, at the current level of experimental precision. Let
us remind a few basic features of this scheme relevant to the
forthcoming discussion of cosmological consequences, in particular,
properties of the DM.

We start with the simplest way to introduce vector-like gauge
interactions of elementary T-quarks and composite T-baryons based
upon the gauged linear $\sigma$-model (GL$\sigma$M)
\cite{Lee,LSigM,SU2LR}. Recently, this scheme was applied to
description of LHC phenomenology of the lowest mass composites --
the physical pseudoscalar T-pions $\tilde{\pi}^{\pm,0}$ and scalar
$\tilde{\sigma}$-meson, as well as possible modifications of the
scalar Higgs boson $h$ couplings, which are relevant for LHC
searches for new strongly-coupled dynamics and precision Higgs
physics \cite{our-TC}.

Consider the local chiral vector-like subgroup $SU(2)_{\rm V\equiv
L+R}$ appearing due to the spontaneous chiral symmetry breaking
(\ref{CSB}) and acting on new confined elementary T-quark and
simultaneously composite T-baryon sectors. For the moment, we do not
assume the condition (\ref{ident}), i.e. we do not explicitly
introduce ordinary weak interactions into the T-quark/T-baryon
sectors. Following to hadron physics analogy, let us extend the
fermion sector by incorporating one Dirac T-nucleon vector-like
doublet $\tilde{N}$ over $SU(2)_{\rm V}$ (an analog of the nucleon
doublet in the SM) in addition to the elementary T-quark vector-like
doublet $\tilde{Q}$ (an analog of the first generation of quarks in
the SM) such that the initial matter fields content of the
vector-like TC model becomes
 \begin{eqnarray} \label{Tdoub}
 \tilde{Q} = \left(
      \begin{array}{c}
         U \\
         D
      \end{array}
             \right)\,, \qquad
  \tilde{N} = \left(
      \begin{array}{c}
         P \\
         N
      \end{array}
             \right)\,,
 \end{eqnarray}
which are in the fundamental representation of the $SU(2)_{\rm
V}\otimes U(1)_{\rm Y}$ group. As usual, in addition we have the
initial scalar T-sigma $S$ field which is the singlet
representation, and the triplet of initial T-pion fields
$P_a,\,a=1,2,3$ which is the adjoint (vector) representation of
$SU(2)_{\rm V}$ (with zeroth $U(1)_{\rm Y}$ hypercharge). Thus, in
terms of the fields introduced above the GL$\sigma$M part of the
Lagrangian responsible for Yukawa-type interactions of the T-quarks
(\ref{Tdoub}) reads
 \begin{eqnarray} \label{Yuk}
 {\cal L}_Y^{\rm TC} = -g^{Q}_{\rm TC} \bar{\tilde{Q}}(S+i\gamma_5\tau_a
 P_a)\tilde{Q}-g^{N}_{\rm TC} \bar{\tilde{N}}(S+i\gamma_5\tau_a
 P_a)\tilde{N}\,, \qquad g^{Q}_{\rm TC}\not=g^{N}_{\rm TC}\,, \quad g^{Q,N}_{\rm TC} > 1 \,,
 \end{eqnarray}
where $\tau_a,\,a=1,2,3$ are the Pauli matrices, and T-strong Yukawa
couplings $g^{Q}_{\rm TC}$ and $g^{N}_{\rm TC}$ are introduced in a
complete analogy to low-energy hadron physics, they absorb yet
unknown non-perturbative strongly-coupled dynamics and can be chosen
to be different. Typically, the perturbativity condition requires
them to be bounded, $g^{Q,N}_{\rm TC}<\sqrt{4\pi}$, in order to
trust predictions of the linear model. After the EWSB phase, the
Yukawa interactions (\ref{Yuk}) will play an important role
determining the strength of T-neutron $N$ self-interactions leading
to specific properties of the associated DM which will be studied
below.

In the SM, the ordinary gauge boson-hadron interactions are usually
introduced by means of gauge bosons hadronisation effects. In the
case of a relatively large T-confinement scale
$\Lambda_{\rm{TC}}\sim 0.1-1$ TeV relevant for our study, the effect
of T-hadronisation of light $W,\,Z$ bosons into heavy composite
states is strongly suppressed by large constituent masses of
T-quarks $M_{Q}\sim \Lambda_{\rm TC}$. Following to arguments of
Ref.~\cite{our-TC}, the vector-like interactions of $\tilde{Q}$,
$\tilde{N}$ and $P_a$ fields with initial $U(1)_{\rm Y}$ and
$SU(2)_{\rm V}$ gauge fields $B_{\mu},\,V_{\mu}^a$, respectively,
can be safely introduced in the local approximation via covariant
derivatives over the local $SU(2)_{\rm V}\otimes U(1)_{\rm Y}$ group
in the same way as ordinary SM gauge interactions, i.e.
 \begin{eqnarray}
  {\cal L}_{\rm kin}^{\rm TC} = \frac12 \partial_{\mu} S\, \partial^{\mu} S
  + \frac12 D_{\mu} P_a\, D^{\mu} P_a +
  i \bar{\tilde{Q}}\hat{D}\tilde{Q} +
  i \bar{\tilde{N}}\hat{D}\tilde{N}\,. \label{LG}
 \end{eqnarray}
Here, covariant derivatives of $\tilde{Q}$, $\tilde{N}$ and $P_a$
fields with respect to $SU(2)_{\rm V}\otimes U(1)_{\rm Y}$
interactions read
 \begin{eqnarray}
    && \hat{D}\tilde{Q} = \gamma^{\mu} \left( \partial_{\mu}
       - \frac{iY_Q}{2}\, g_1B_{\mu} - \frac{i}{2}\, g^V_2 V_{\mu}^a \tau_a
       \right)\tilde{Q}\,, \nonumber \\
    && \hat{D}\tilde{N} = \gamma^{\mu} \left( \partial_{\mu}
       - \frac{iY_N}{2}\, g_1B_{\mu} - \frac{i}{2}\, g^V_2 V_{\mu}^a \tau_a
       \right)\tilde{N}\,, \label{DQ} \\
    && \qquad D_{\mu} P_a = \partial_{\mu} P_a + g^V_2
\epsilon_{abc} V^b_{\mu} P_c\,, \nonumber
 \end{eqnarray}
respectively, besides that $\tilde{Q}$ is also assumed to be
confined under a QCD-like $SU(n)_{\rm TC}$ group. Below, for the
sake of simplicity we discuss a particular case with the number of
T-colors $n=3$, analyze a possible implementation of EW interactions
into T-quark/T-baryon sectors according to
\begin{eqnarray}
SU(2)_{\rm V}\to SU(2)_{\rm W}\,, \qquad V_{\mu}^a\to W_{\mu}^a\,,
\qquad g^V_2\to g_2 \label{ident-1}
\end{eqnarray}
replacement rule in Eq.~(\ref{DQ}). A consistency test of the latter
scenario against the DM relic abundance and direct DM detection data
for rank-2 confined group will enable us to draw important
conclusions about properties of TC sectors.

The additional gauge and Yukawa parts (\ref{Yuk}) and (\ref{LG})
should be added to the SM Lagrangian written in terms of SM gauge
$B_{\mu},\,W_{\mu}^a$ and chiral fields as follows
\begin{eqnarray}
 \begin{array}{c} \vspace*{5mm}
  \displaystyle {\cal L}^{\rm gauge}_{\rm SM} = -\frac{1}{2}g_1B_\mu\bar l_L\gamma^\mu l_L -
  g_1B_\mu\bar e_R\gamma^\mu e_R + \frac{1}{6}g_1B_\mu\bar q_L\gamma^\mu q_L +
  \frac{2}{3}g_1B_\mu\bar u_R\gamma^\mu u_R - \frac{1}{3}g_1B_\mu\bar d_R\gamma^\mu d_R
\\
  \displaystyle
  + \frac{1}{2}g_2W^a_\mu\bar l_L\gamma^\mu\tau_a l_L +
    \frac{1}{2}g_2W^a_\mu\bar q_L\gamma^\mu\tau_a q_L\,.
 \end{array}
 \label{LSM1}
\end{eqnarray}
Here summation over flavor and family indices is implied. The theory
in its simplest formulation discussed here, of course, does not
predict particular values for elementary and composite T-quark
hypercharges $Y_Q$ and $Y_N$. These, together with the number of
T-quark generations, the respective properties of interactions, the
group of confinement, etc. should be ultimately constrained in
extended chiral-gauge or grand-unified theories along with coming
experimental data. Employing further analogies with the SM and QCD,
in what follows we fix the hypercharge of the elementary T-quark
doublet to be the same as that of quark doublets in the SM, i.e.
$Y_{Q}=1/3$, and the hypercharge of the T-nucleon doublet -- to be
the same as that of nucleon doublet in the SM, i.e. $Y_{N}=1$. Thus,
the T-baryon states in Eq.~(\ref{Tdoub}) become T-nucleons composed
of three elementary T-quarks, i.e. $P=(UUD)$, $N=(DDU)$ in analogy
to proton and neutron in QCD. Other assignments with different
hypercharges and number of T-colors are also possible and would lead
to other possible types of T-baryons. The basic qualitative results
for the DM properties in odd $SU(3)_{\rm TC}$ confined group of
QCD-type are generic for other odd $SU(2n+1)_{\rm TC},n=2,3,\dots$
groups. In this work we stick to a direct analogy with QCD for
simplicity and test it against available DM constraints.

One of the interesting but alternative opportunities would be to
consider $Y_Q=0$ case such that an integer electric charge of
T-baryons would only be possible for even TC groups $SU(2n)_{\rm
TC}$ with the simplest $SU(2)_{\rm TC}$. Here, T-baryons are
two-T-quark systems. In the non-perturbative T-hadron vacuum the
$UD$ state with zeroth electric charge is energetically favorable
since extra binding energy appears due to exchanges of collective
excitations with T-pion quantum numbers (in usual hadron physics the
effect of $ud$-coupling brings up extra $70$ MeV into the binding
energy) making the neutral di-T-quark $UD$ state to be absolutely
stable and thus an appealing DM candidate. This case has certain
advantages and will be considered elsewhere.

After $SU(2)_{\rm V}\otimes U(1)_{\rm Y}$ symmetry breaking an extra
set of heavy gauge $Z',{W'}^\pm$ bosons interacting with T-quark and
T-baryons emerges\footnote{Such extra $Z',{W'}^\pm$ bosons can, in
principle, be composite and identified with composite $\rho^{0,\pm}$
mesons or elementary vector bosons from ``right isospin''
$SU(2)_{\rm R}$ group as a part of chiral-symmetric $SU(2)_{\rm
L}\otimes SU(2)_{\rm R}$ extension of the SM. This point, however,
is not critical for the current study of the QCD-type T-neutron DM
properties and we do not discuss it here.}. If one does not imply a
straightforward identification (\ref{ident-1}), T-quarks and
T-baryons may still interact with ordinary SM gauge fields via a
(very small) mixing between $Z'$ and $Z$, ${W'}^\pm$ and $W^\pm$
bosons, respectively. Of course, such a mixing must be tiny to not
spoil the EW precision tests. An alternative option would be to
adopt (\ref{ident-1}) where the EW precision tests are satisfied
without any serious tension \cite{our-TC}. Then, in analogy to
constituent colored T-quarks \cite{our-TC}, the vector-like SM gauge
interactions of T-baryons with $Z,W^\pm$ bosons are controlled by
the following part of the Lagrangian
\begin{eqnarray}
 && L_{\bar{\tilde{N}}\tilde{N}Z/W}=\delta_W\frac{g_2}{\sqrt{2}}\,\bar{P}\gamma^{\mu}N
 \cdot W_{\mu}^+ + \delta_W\frac{g_2}{\sqrt{2}}\,\bar{N}\gamma^{\mu}P\cdot
 W_{\mu}^- \nonumber \\
 && \quad +\, \delta_Z\frac{g_2}{c_W}\, Z_{\mu} \sum_{f=P,N}
\bar{f}\gamma^{\mu}\bigl(t_3^f-q_f\,s^2_W\bigr)f\,. \label{L-QV}
\end{eqnarray}
Here, $\delta_{W,Z}$ are the generic parameters which control EW
interactions of T-baryons, $e=g_2 s_W$ is the electron charge,
$t_3^f$ is the weak isospin ($t_3^P=1/2$, $t_3^N=-1/2$),
$q_f=Y_N/2+t_3^f$ is the T-baryon charge. The two consistent options
for introducing weak interactions into the T-fermion sectors
dictated by EW precision tests can be summarized as follows:
\begin{eqnarray} \nonumber
 && {\rm I.} \qquad\; \delta_{W,Z}=1\,, \qquad \qquad  \; SU(2)_{\rm V}\simeq SU(2)_{\rm
 W}\,, \\
 && {\rm II.} \qquad \delta_{W,Z}\ll 1\,, \qquad \qquad SU(2)_{\rm
 V}\not= SU(2)_{\rm W}\,, \quad m_{Z',W'}\gg 100\, {\rm GeV}\,.
 \label{scenarios}
\end{eqnarray}
In the first case, one deals with pure EW vector interactions of
T-quarks/T-baryons corresponding to transition (\ref{ident-1}),
while in the second case $\delta_{W,Z}$ are related to a very small
mixing between SM vector bosons and extra {\it different}
$SU(2)_{\rm V}$ bosons tagged as $Z'$ and ${W'}^\pm$. In both cases,
couplings with photons are not (noticeably) changed and are
irrelevant for Dirac T-neutron DM studies, so are not shown here. We
will test the both options above against available constraints on
T-neutron DM implying the existence of $SU(3)_{\rm TC}$ group in
confinement.

As agreed above, we choose $Y_N=1$ in analogy to the SM, thus
$q_P=1$ and $q_N=0$ as anticipated. The Yukawa-type interactions of
T-baryons with scalar ($h$ and $\tilde{\sigma}$) and pseudoscalar
($\tilde{\pi}^{0,\pm}$) fields are driven by
\begin{eqnarray}
 &&L_{\bar{\tilde{N}}\tilde{N}h}+L_{\bar{\tilde{N}}\tilde{N}\tilde{\sigma}}+
 L_{\bar{\tilde{N}}\tilde{N}\tilde{\pi}}=-g^{N}_{\rm
TC}\,(c_{\theta}\tilde{\sigma}+s_{\theta}h)\cdot (\bar{P}P +
\bar{N}N) \nonumber \\ && - i\sqrt{2}g^{N}_{\rm
TC}\,\tilde{\pi}^+\bar{P}\gamma_5 N - i\sqrt{2}g^N_{\rm
TC}\,\tilde{\pi}^-\bar{N}\gamma_5 P - ig^N_{\rm TC}\,
\tilde{\pi}^0(\bar{P}\gamma_5 P - \bar{N}\gamma_5 N)\,.
\label{L-QhSpi}
\end{eqnarray}
The gauge and Yukawa parts of the Lagrangian (\ref{L-QV}) and
(\ref{L-QhSpi}) completely determine the T-baryon interactions at
relatively low kinetic energies $E_{\rm kin}\ll M_{B_{\rm T}}$
typical for equilibrium reactions (scattering, production and
annihilation) processes in the cosmological plasma before DM thermal
freeze-out epoch (see below). Note that due to vector-like nature of
extra virtual T-baryon states they do not produce any noticeable
contributions to the oblique corrections and FCNC processes
preserving internal consistency of the model under consideration
\cite{our-TC}. The latter is true for both models I and II
(\ref{scenarios}).

The interactions of T-pions with $Z,\,W^\pm$ bosons which will be
used in further calculations of T-baryon annihilation cross sections
are defined as follows
\begin{eqnarray} \nonumber
 L_{\tilde{\pi}\tilde{\pi}Z/W} &=&
 i g_2^W {W^{\mu}}^+ \cdot (\tilde{\pi}^0\tilde{\pi}_{,\mu}^- -
 \tilde{\pi}^-\tilde{\pi}_{,\mu}^0) + i g_2^W {W^{\mu}}^- \cdot
 (\tilde{\pi}^+\tilde{\pi}_{,\mu}^0 -
 \tilde{\pi}^0\tilde{\pi}_{,\mu}^+) \\
 &+& i g^Z_2 c_W Z_{\mu}\cdot
 (\tilde{\pi}^-\tilde{\pi}_{,\mu}^+ -
 \tilde{\pi}^+\tilde{\pi}_{,\mu}^-)\,,  \label{L-piV}
\end{eqnarray}
where $g_2^{W,Z}=g_2\,\delta_{W,Z}$, $\tilde{\pi}_{,\mu} \equiv
\partial_{\mu}\tilde{\pi}$. The Yukawa interactions
$\bar{f}fh + \bar{f}f\tilde{\sigma}$ of the ordinary fermions get
modified compared to the SM as follows
\begin{eqnarray}
L_{\bar{f}fh}+L_{\bar{f}f\tilde{\sigma}}=-g_2(c_{\theta}h-s_{\theta}\tilde{\sigma})\cdot
\frac{m_f}{2M_W}\bar{f}f\,. \label{Yukawa}
\end{eqnarray}
The Lagrangians of the $h\tilde{\pi}\tilde{\pi}$ and $hWW+hZZ$
interactions will also be needed below, so we write them down here
as well:
\begin{eqnarray} \nonumber
 && L_{h\tilde{\pi}\tilde{\pi}}=-(\lambda_{\rm TC}u\,s_\theta-\lambda
vc_\theta)\,h(\tilde{\pi}^0\tilde{\pi}^0 +
2\tilde{\pi}^+\tilde{\pi}^-)=
-\frac{M_h^2-m_{\tilde{\pi}}^2}{2M_Q}\,g_{\rm TC}s_\theta\,
h(\tilde{\pi}^0\tilde{\pi}^0 + 2\,\tilde{\pi}^+\tilde{\pi}^-)\ ,
\\
 && L_{hWW}+L_{hZZ}=g_2\,M_W c_\theta \,hW_\mu^+{W^{\mu}}^- +
\frac12(g_1^2+g_2^2)^{1/2}M_Zc_\theta\, hZ_\mu Z^\mu\ .
 \label{L-hVVpipi}
 \end{eqnarray}
Finally, the interactions $\tilde{\sigma} \tilde{\pi}\tilde{\pi}$
and $\tilde{\sigma} WW+\tilde{\sigma} ZZ$ are determined by
\begin{eqnarray} \nonumber
 && L_{\tilde{\sigma} \tilde{\pi}\tilde{\pi}}=-(\lambda_{\rm TC}u
c_\theta+\lambda v s_\theta)\,\tilde{\sigma}
(\tilde{\pi}^0\tilde{\pi}^0 + 2\,\tilde{\pi}^+\tilde{\pi}^-)=
-\frac{M_{\tilde \sigma}^2-m_{\tilde{\pi}}^2}{2M_Q}\,g_{\rm
TC}c_\theta\, \tilde{\sigma} (\tilde{\pi}^0\tilde{\pi}^0 +
2\,\tilde{\pi}^+\tilde{\pi}^-)\ ,
\\
 && L_{\tilde{\sigma} WW}+L_{\tilde{\sigma} ZZ}=-g_2\,M_W s_\theta\,
\tilde{\sigma} W_\mu^+{W^{\mu}}^--\frac12(g_1^2+g_2^2)^{1/2}M_Z
s_\theta\, \tilde{\sigma} Z_\mu Z^\mu\ .
 \label{L-SVVpipi}
 \end{eqnarray}

In the considering vector-like TC model, the T-baryon mass scale
$\sqrt{s}\lesssim M_{B_{\rm T}}$ should be considered as an upper
cut-off of the considering model which contains only the lightest
physical d.o.f. $\tilde \pi$ and $\tilde \sigma$. The latter are
sufficient in the current first analysis of the vector-like T-baryon
DM in the non-relativistic limit $v_B\ll 1$. Certainly, at higher
energies $\sqrt{s}\gtrsim M_{B_{\rm T}}$ the theory should involve
higher (pseudo)vector and pseudoscalar states (e.g. $\tilde \rho$,
$\tilde a_0$, $\tilde a_1$ etc). The latter extension of the model
will be done elsewhere if required by phenomenology.

\section{T-baryon mass spectrum}

Typical (but optional) assumption about a dynamical similarity
between color and T-color in the case of confined $SU(3)_{\rm TC}$
enables us to estimate characteristic masses of the lightest
T-hadrons and constituent T-quarks through the scale transformation
of ordinary hadron states via an approximate scale factor
$\zeta=\Lambda_{\rm TC}/\Lambda_{\rm QCD}\gtrsim 1000$ following
from a relative proximity of EW scale and $\Lambda_{\rm
TC}\sim0.1-1$ TeV, i.e.
 \begin{eqnarray}
m_{\tilde{\pi}}\gtrsim 140\,{\rm GeV}\,,\quad
M_{\tilde{\sigma}}\gtrsim 500\,{\rm GeV}\,,\quad M_{Q}\gtrsim
300\,{\rm GeV}\,,\quad M_{B_{\rm T}}\equiv M_{P}\simeq M_{N}\gtrsim
1\,{\rm TeV}\,, \label{QCD-an}
 \end{eqnarray}
for the T-pion $m_{\tilde{\pi}}$, T-sigma $M_{\tilde{\sigma}}$ and
constituent T-quark $M_{Q}$ and T-baryon $M_{B_{\rm T}}$ mass
scales. In QCD, the constituent quark masses roughly take a third of
the nucleon mass, so it is reasonable to assume that the same
relation holds in T-baryon spectrum
 \begin{eqnarray}
  M_{B_{\rm T}}\equiv M_P\simeq M_N \simeq 3M_Q=3g^Q_{\rm TC}u\,,
 \end{eqnarray}
where $u\sim\Lambda_{\rm TC}=0.1-1$ TeV is the T-sigma vacuum
expectation value (vev) which spontaneously breaks the local chiral
symmetry in the T-quark sector down to weak isospin group
(\ref{CSB}) (for more details on the chiral and EW symmetries
breaking in the considering model, see Ref.~\cite{our-TC}). In the
chiral limit of the theory, T-sigma vev $u$ has the same
quantum-topological nature as the SM Higgs vev $v\simeq 246$ GeV,
i.e. $u,\, v \sim |\langle\bar{\tilde{Q}}\tilde{Q}\rangle|^{1/3}$ in
terms of the T-quark condensate
$|\langle\bar{\tilde{Q}}\tilde{Q}\rangle|\not=0$ providing the
dynamical nature of the EWSB mechanism in the SM.

Also, with respect to interactions with known particles at typical
4-momentum squared transfers $Q^2\ll l_{\rm TC}^{-2}\gtrsim
2.3\,{\rm TeV}^2$, where $l_{\rm TC}$ is the characteristic length
scale of the non-perturbative T-gluon fluctuations estimated by
rescaling of that from QCD (\ref{QCD-an}), the T-hadrons behave as
elementary particles with respect to EW interactions. Besides DM
astrophysics, not very heavy vector-like T-baryons can also be
relevant for the LHC phenomenology as well which is an important
subject for further studies.

Adopting the hypothesis about T-baryon number conservation in
analogy to ordinary baryon number, let us find constraints on the
vector-like TC model parameters providing an inverse mass hierarchy
between T-neutron and T-proton, i.e. $M_{N}<M_{P}$. In this case,
T-neutron becomes indeed the lightest T-baryon state and, hence,
stable which makes it an appealing DM candidate.

In usual hadron physics it is known that the isospin $SU(2)$
symmetry at the level of current quark masses is strongly broken --
the current mass difference between $u$ and $d$ quarks is of the
order of their masses. Such a symmetry is restored to a good
accuracy at the level of constituent quarks and nucleons. This
restoration is a direct consequence of smallness of the current
quark masses compared to contributions from the non-perturbative
quark-gluon vacuum to the hadron masses. A small mass splitting in
the hadron physics is typically estimated in the baryon-meson theory
which operates with hadron-induced corrections (in particular,
$\rho$-meson loops with a $\rho$-$\gamma$ mixing).

In the case of the local vector-like subgroup $SU(2)_{\rm V}$ in
both models I and II we neglect T-rho $\tilde \rho$ mediated
contributions to respective DM annihilation cross sections assuming
for simplicity that $\tilde \rho$ mixing with $\gamma$ and $Z$ is
very small due to a strong mass hierarchy between them. In this
simplified approach we can evaluate the {\it lower bound} on the
T-baryon mass splitting induced by pure EW corrections only (other
EW-like gauge interactions and non-local effects may only increase
it). The T-strong interactions do not distinguish isotopic
components in the T-baryon doublet, and thus do not contribute to
the mass splitting between $P$ and $N$.

When the loop momentum becomes comparable to the T-baryon mass scale
$M_{B_{\rm T}}\gg m_Z$ or higher, a T-baryon parton substructure
starts to play an important role. In particular, the local
approximation for EW T-baryon interactions does not work any longer
and one has to introduce non-local Pauli form factors instead of the
local gauge couplings. The latter must, in particular, account for
non-zeroth anomalous magnetic moments of T-baryons. In the EW loop
corrections to the T-baryon mass splitting, however, typical
momentum transfers $q$ which dominate the corresponding (finite)
Feynman integral are at the EW scale $M_{\rm EW}\sim 100$ GeV scale.
At such scales we can safely neglect non-local effects and use
ordinary local gauge couplings renormalized at $\mu^2=M_{B_{\rm
T}}^2$ scale. The latter approximation is sufficient for a rough
estimate of the mass splitting and, most importantly, its sign.

Note the coincidence of T-isotopic $SU(2)_{\rm V}$ symmetry at the
fundamental T-quark level with the weak isospin $SU(2)_{\rm W}$ of
the SM provides arbitrary but {\it exactly equal} current T-quark
masses in the initial Lagrangian, $m_U=m_D$, such that fundamental
T-quark and hence T-baryon spectra are degenerate at tree level.
Note in the initial SM Lagrangian current $u,d$-quark masses are
equal to zero due to chiral asymmetry of weak interactions. After
spontaneous EW symmetry breaking very different current $u,d$-quark
masses emerge as a consequence of the absence of $SU(2)$
interactions for right-handed $u,d$ quarks.

\begin{figure}[h!]
 \centerline{\includegraphics[width=0.95\textwidth]{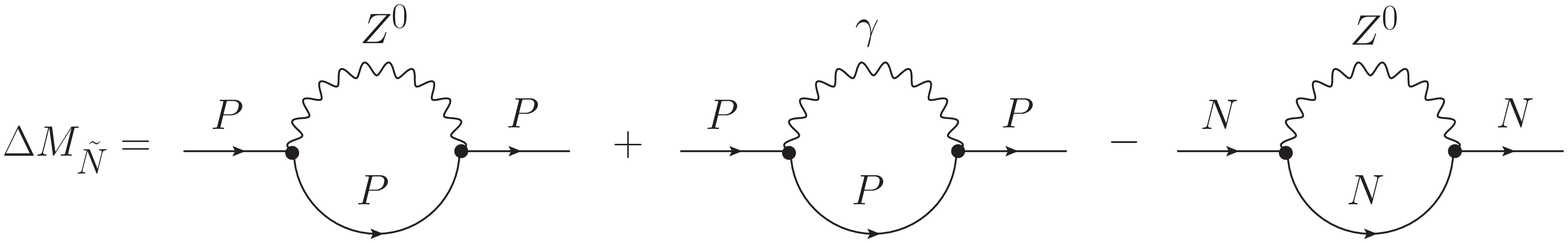}}
   \caption{\small One-loop EW radiative corrections causing positive
   mass splitting between T-proton and T-neutron in
   the chiral limit of the underlined theory, $\Delta M_{B_{\rm T}}^{\rm EW}>0$.
   Other corrections from $W^{\pm}$, T-pion, T-sigma and Higgs boson loops
   enter symmetrically to $P$ and $N$ self-energies and thus do
   not contribute to the mass splitting $\Delta M_{B_{\rm T}}^{\rm EW}$ and
   not shown here.}
   \label{fig:split}
\end{figure}

In the considering case of degenerate vector-like T-quark mass
spectrum $\Delta M_Q\equiv M_U-M_D=0$ the EW radiative corrections
dominate the mass splitting between T-proton and T-neutron for
suppressed heavy T-rho $\tilde \rho$ contributions and a small
$\tilde \rho$-gauge bosons mixing, $\Delta M_{B_{\rm T}}^{\rm
EW}\lesssim\Delta M_{B_{\rm T}}\equiv M_P-M_N\ll M_{B_{\rm T}}$.
Corresponding EW ($Z,\gamma$-mediated) one-loop diagrams are shown
in Fig.~\ref{fig:split}.

In the model I (\ref{scenarios}) the T-baryon mass splitting $\Delta
M_{B_{\rm T}}^{\rm EW}$ due to EW corrections is given in terms of
the difference between the T-baryon mass operators on mass shell
which takes a form of the following finite integral
\begin{eqnarray}
 \Delta M_{B_{\rm T}}^{\rm EW}=-\frac{ie^2M_{Z}^2}{8\pi^4}\int\frac{(\hat{q}-M_{B_{\rm T}})dq}
 {q^2(q^2-M_{Z}^2)[(q+p)^2-M_{B_{\rm T}}^2]}\,,
\end{eqnarray}
given by $\gamma$ and $Z$ corrections shown in Fig.~\ref{fig:split}
only. Note that logarithmic divergences explicitly cancel out in the
difference between $P$ and $N$ mass operators, providing us with the
finite result for $\Delta M_{B_{\rm T}}^{\rm EW}$. Other corrections
from $W^{\pm}$, scalar and pseudoscalar loops enter symmetrically
into $P$ and $N$ self-energies and thus are canceled out too. In the
realistic case of heavy T-baryons $M_{B_{\rm T}}\gg M_Z$ we arrive
at the following simple relation
\begin{eqnarray}
  \Delta M_{B_{\rm T}}^{\rm EW} \simeq \frac{\alpha(M_{B_{\rm T}}) M_Z}{2}>0\,,
 \label{deltaM-eq}
\end{eqnarray}
where the fine structure constant $\alpha(\mu)$ is fixed at the
T-baryon scale $\mu=M_{B_{\rm T}}\sim 1$ TeV. Numerically, we find
to a good accuracy
 \begin{eqnarray}
 \Delta M_{B_{\rm T}}^{\rm EW} \simeq 360\; {\rm MeV}\,. \label{deltaM-num}
 \end{eqnarray}
This can be considered as the EW contribution to the T-baryon mass
splitting and provides a conservative lower limit to it. The
T-proton is not stable and weakly decays into T-neutron and light SM
fermions ($e,\,\mu,\,\nu_{e,\mu},\,u,\,d,\,s$) as follows $P\to
N+(W^*\to f_i\bar{f}_j)$. Remarkably enough, the EW radiative
corrections appear to work in the right direction making the
T-proton slightly heavier than the T-neutron such that the latter
turns out to be stable and viable as a heavy DM candidate. With the
mass splitting value (\ref{deltaM-num}), we find the following
approximate vector-like T-proton lifetime
\begin{eqnarray}
 \tau_P \simeq \frac{15\pi^3}{G_F^2}(\Delta M_{B_{\rm T}}^{\rm EW})^{-5}
 \simeq 0.4 \times 10^{-9}\;{\rm s}\,.
 \label{P-life}
\end{eqnarray}

In the model II, the $Z$ contributions die out in the limit
$\delta_{\rm EW}\ll 1$, so it can only be induced by extra heavy
$Z'$ exchange. The corresponding contribution to the $P$-$N$ mass
splitting $\Delta M_{B_{\rm T}}^{\rm V}$ is obtained from
Eq.~(\ref{deltaM-eq}) by a replacement $m_Z\to m_Z'$,
\begin{eqnarray}
  \Delta M_{B_{\rm T}}^{\rm V} \simeq \frac{\alpha(M_{B_{\rm T}})
  M_{Z'}}{2}>0\,, \qquad M_{Z'} < M_{B_{\rm T}}\,,
 \label{deltaM-eq-V}
\end{eqnarray}
such that the $\Delta M_{B_{\rm T}}^{\rm V}\gg \Delta M_{B_{\rm
T}}^{\rm EW}$ and the T-proton lifetime would even be shorter. Of
course, the estimate (\ref{deltaM-eq-V}) should be taken with care
for $m_{Z'}\gtrsim M_{B_{\rm T}}$ when non-local effects become
important, and a radiative mass splitting between constituent $U$
and $D$ T-quarks would determine the actual mass difference between
$P$ and $N$.

Note, in the most natural and simplest model I the properties of the
vector-like T-baryon spectrum are very similar to properties of
vector-like Higgsino LSP (e.g. splitting between chargino and
neutralino) spectrum due to practically the same structure of EW
interactions \cite{our-Higgsino}. The key difference between the
lightest Higgsino and T-neutron DM candidates is in capability of
T-neutrons to self-interactions (e.g. enhanced self-annihilation and
elastic scattering rates) driven essentially by T-strong Yukawa
terms (\ref{L-QhSpi}) which make them specifically interesting for
DM phenomenology and astrophysics.

\section{Annihilation of T-baryons in cosmological plasma}

In the considering vector-like TC model the interaction properties
of T-baryons are fixed by gauge and Yukawa interactions determined
by Eqs.~(\ref{L-QV}) and (\ref{L-QhSpi}), while the radiative
splitting in the mass spectrum between $P$ and $N$ states is given
by Eq.~(\ref{deltaM-eq}) or (\ref{deltaM-eq-V}). Besides an unknown
rank of the confined group, physics of vector-like T-baryon DM, its
interaction properties and formation depend quantitatively on four
{\it physical} parameters only: the strong T-hadron coupling $g_{\rm
TC}^N$, T-quark mass scale $M_Q$, the T-pion $m_{\tilde \pi}$ and
T-sigma $M_{\tilde \sigma}$ masses. The physically interesting
parameter space domain corresponds to relatively small ${\tilde
\sigma}h$-mixing angle $s_\theta\lesssim 0.2$ where the oblique
corrections (essentially, $T$-parameter) are small and deviations
from the SM Higgs couplings are strongly suppressed and can be well
within the current LHC constraints (see Ref.~\cite{our-TC} and
references therein). Assuming that the DM consists of composite
T-baryons (mostly, T-neutrons with probably a small fraction of
anti-T-neutrons), let us discuss extra possible constraints on the
vector-like TC parameter space coming from astrophysical
observations and cosmological evolution of the DM (for a review on
the current (in)direct DM detection measurements and constraints,
see e.g.
Refs.~\cite{Schnee:2011ef,Cirelli:2012tf,Drees:2012ji,Bergstrom:2012fi}).

In order to estimate the T-baryon mass scale $M_{B_{\rm T}}$ from
the DM relic abundance data \cite{Planck} one has to consider
evolution of the T-baryon density in early Universe which is largely
determined by the kinetic $B_{\rm T}{\bar B}_{\rm T}$ annihilation
cross section $(\sigma v_B)_{\rm ann}$. As is typical for the cold
DM formation scenarios one naturally assumes that the residual
T-baryon abundance is formed at temperatures $T\ll M_{B_{\rm T}}$
when non-relativistic approximation is applied.

In practice, the phenomenologically acceptable domain of the
vector-like TC parameter space corresponding to $s_\theta\ll 1$
means that the hierarchy of EW and chiral symmetry breaking scales
is far from degeneracy, i.e. $u/v\gg 1$. This means that the
realistic T-baryon mass scale can be decoupled, although not very
strongly, from the EW one, i.e. $M_{B_{\rm T}}\gg M_{\rm EW}$ for
not too small T-quark/T-baryon Yukawa couplings $g_{\rm
TC}^{N,Q}\gtrsim 1$. The latter means that the T-baryon spectrum,
and possibly (pseudo)scalar spectrum, would be way above the Higgs
boson mass scale, and the naive QCD scaling (\ref{QCD-an}) can be
satisfied.

On the other hand, it is also possible that T-pions evade LEP II and
current low mass LHC constraints due to very small (T-quark loop
induced) production cross sections and narrow widths, and thus they
can, in principle, be as light as $W$ boson, $m_{\tilde \pi}\gtrsim
m_W$. While T-sigma is extremely wide $\Gamma_{\tilde \sigma}\sim
M_{\tilde \sigma}$ and hard to be detected with current collider
techniques, it would be possible for it to have a rather low mass
down to $\sim 150$ GeV or even lower without upsetting current EW
precision and LHC constraints. Note, current LHC constraints on the
T-pion mass from the ordinary TC (e.g. Extended TC) scenarios do not
apply to the considering vector-like TC model where T-pions do not
couple to ordinary fermions.

The last two possibilities of decoupled and non-decoupled TC sectors
in Nature make it reasonable to consider the irreversible
annihilation of T-baryons in two different phases of cosmological
plasma separately -- before and after EW phase transition epoch
$T_{\rm EW}\sim 200$ GeV. Consequently, we will end up with two
different scenarios of the DM relic abundance formation which have
to be (qualitatively) discussed in detail.

\subsection{Annihilation of vector-like T-baryons: the high-symmetry phase}

At temperatures $T>T_{\rm EW}\sim 200$ GeV the Higgs condensate
$\langle H\rangle \equiv v$ is melted, i.e. $v=0$, and thus weak
isospin $SU(2)_{\rm W}$ of the SM is restored, while the T-sigma
condensate does not vanish $\langle S\rangle \equiv u\not=0$, $u\gg
T_{\rm EW}$, such that the chiral symmetry in the fundamental
T-quark sector is broken: $SU(2)_{\rm L}\otimes SU(2)_{\rm R}\to
SU(2)_{\rm V}$. In what follows, we refer to this period in the
cosmological evolution as to {\it the high-symmetry (HS) phase of
the cosmological plasma} with the characteristic temperature $T_{\rm
EW}<T\lesssim u$. This means that the T-baryon mass scale should be
well above the EW scale for the DM relic abundance to be formed
entirely in the HS phase, i.e. $M_{B_{\rm T}}\gg 200$ GeV.
\begin{figure}[t!]
 \centerline{\includegraphics[width=0.9\textwidth]{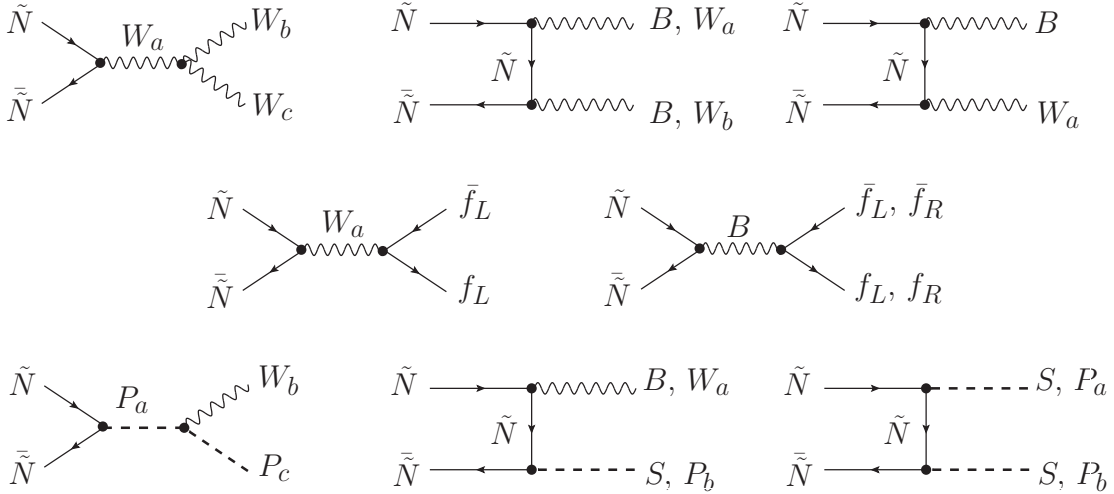}}
   \caption{\small Typical diagrams contributing to the T-baryon
   DM annihilation in the high-symmetry phase of the cosmological
   plasma corresponding to $T_{\rm EW}<T\lesssim u$. The model I
   (\ref{scenarios}) is implied.}
   \label{fig:HS-ann}
\end{figure}

In the HS case with $SU(2)_{\rm V}=SU(2)_{\rm W}$ (model I), the
equilibrium number densities of (anti)T-neutrons and (anti)T-protons
are equal to each other $n_N = n_P$ ($n_{\bar N} = n_{\bar P}$)
since the T-baryon mass spectrum is degenerate, i.e. $\Delta
M_{B_{\rm T}}=0$ (or more precisely, $T\gg \Delta M_{B_{\rm T}}$),
so the total T-baryon number density is $n_{B_{\rm T}}\simeq
2(n_N+n_{\bar N})$, at least, before the T-baryon freeze out epoch.
Practically, $P$ and $N$ states are dynamically equivalent in this
phase and participate in all reactions as components of the isospin
$SU(2)_{\rm V}$ doublet ${\tilde N}$ (\ref{Tdoub}) with Yukawa and
gauge interactions determined by Eqs.~(\ref{Yuk}) and (\ref{LG}),
respectively. Consequently, masses of all SM fermions and gauge
bosons vanish in this phase (more precisely, $m_f,\,M_{W,Z} \ll T$),
while T-sigma $M_{\tilde \sigma}$ and T-pion $m_{\tilde \pi}$ masses
(related as $M_{\tilde \sigma}\simeq \sqrt{3}m_{\tilde \pi}$ in the
limiting ``no $\tilde{\sigma}h$-mixing'' case) do not vanish but are
likely to be much smaller than the T-baryon mass scale $M_{B_{\rm
T}}\sim u$ since $u\gg v$, i.e. $M_{\tilde \sigma},\,m_{\tilde
\pi}\ll M_{B_{\rm T}}$. Thus, all the masses except for T-baryon
mass can be neglected in practical calculations to a good
approximation. So, in the HS phase we effectively end up with the
single T-baryon mass scale parameter $M_{B_{\rm T}}$, which has to
be constrained together with the strong Yukawa coupling $g_{\rm
TC}^N$ from astrophysics data.

Let us evaluate the vector-like T-baryon annihilation cross section
$(\sigma v_B)_{\rm ann}$ in the HS phase of the cosmological plasma
in the model I (\ref{scenarios}). All relevant contributions are
schematically depicted in Fig.~\ref{fig:HS-ann}. In comparison with
the Higgsino LSP scenario in $SU(5)$ Split SUSY Model
\cite{our-Higgsino}, the T-baryon annihilation in the HS phase is
given by essentially the same EW amplitudes due to the same
vector-like structure of T-baryon and Higgsino EW interactions, i.e.
\begin{eqnarray} \nonumber
 && \qquad\qquad\qquad {\tilde N}\bar{{\tilde N}} \to BB\,,\qquad
 {\tilde N}\bar{{\tilde N}} \to BW_a\,,\qquad
 {\tilde N}\bar{{\tilde N}} \to  W_aW_b\,,
\\
 && {\tilde N}\bar{{\tilde N}} \to B^* \to
 l_L\bar l_L\,,\;q_L\bar q_L\,,\;e_R\bar e_R\,,\;
 u_R\bar u_R\,,\;d_R\bar d_R\,,\qquad
 {\tilde N}\bar{{\tilde N}} \to W_a^* \to
 l_L\bar l_L,\;q_L\bar q_L\,,
 \label{HS-EW}
\end{eqnarray}
where $l_L,\;q_L$, and $e_R,\;u_R,\;d_R$ are the $SU(2)_W$ doublet
and singlet (chiral) leptons and quarks, respectively, in each of
three generations. The corresponding EW contribution to the total
T-baryon annihilation cross section in the HS phase for
non-relativistic T-baryons $v_B\ll 1$ is found to be
\begin{eqnarray}
 \displaystyle (\sigma v_B)_{\rm ann}^{\rm EW}=
 \frac{21g_1^4+6g_1^2g_2^2+39g_2^4}{512 \pi M_{B_{\rm T}}^2}\,.
 \label{sHS-EW}
\end{eqnarray}
Here $g_1=g_1(\sqrt{s}),\,g_2=g_2(\sqrt{s})$ are the EW gauge
couplings fixed at the scale $\sqrt{s}\simeq2M_{B_{\rm T}}$.

In addition to the pure EW channels listed above, there are a few
important T-strong channels with primary T-pion $P_a$ and T-sigma
$S$ in intermediate and final states. In particular, the
annihilation channels into a (pseudo)scalar and a massless gauge
boson involving additional Yukawa interactions in the T-hadron
sector are
\begin{eqnarray}
 {\tilde N}\bar{{\tilde N}} \to P_a^* \to P_bW_c\,,\qquad
 {\tilde N}\bar{{\tilde N}} \to P_aB\,,\;SB\,,\;SW_a\,,
 \label{HS-TC-1}
\end{eqnarray}
and the corresponding total cross section in the limit $M_{B_{\rm
T}}\gg M_{\tilde \sigma},\,m_{\tilde \pi}$ is
\begin{eqnarray}
 (\sigma v_B)_{\rm ann}^{\rm EW+TC}\simeq
 \frac{(g_{\rm TC}^N)^2(2g_1^2+3g_2^2)}{32\pi M_{B_{\rm T}}^2} \,.
 \label{HS-TC-1-CS}
\end{eqnarray}
In order to turn to the model II (\ref{scenarios}) in the limit
$m_{Z'}\ll M_{B_{\rm T}}$ corresponding to unbroken $SU(2)_{\rm
V}\not=SU(2)_{\rm W}$, one has to perform a replacement $g_2\to
g_2^{\rm V}$ in Eqs.~(\ref{sHS-EW}) and (\ref{HS-TC-1-CS}). This
would provide a rough estimate for the annihilation cross sections
into $B_{\mu}, V^a_{\mu}$ bosons. For a more precise analysis of the
gauge $SU(2)_{\rm V}$ part of the cross sections one should consider
details of the broken phase of $SU(2)_{\rm V}$ and evaluate them for
massive $Z',{W'}^\pm$ bosons for kinematically allowed channels,
i.e. for $m_{W',Z'}\lesssim M_{B_{\rm T}}/2$ (for more details, see
calculations in the low-symmetry phase below). The latter, however,
do not affect our conclusions here since corresponding cross
sections are relatively small compared to those in the T-strong
channels.

For pure T-strong channels
\begin{eqnarray}
 {\tilde N}\bar{{\tilde N}} \to P_aP_b \,,\; SP_a \,,\; SS \,,
 \label{HS-TC-2}
\end{eqnarray}
we have the total cross section
\begin{eqnarray}
 (\sigma v_B)_{\rm ann}^{\rm TC} \simeq
 \frac{9(g_{\rm TC}^N)^4}{32\pi M_{B_{\rm T}}^2}\,.
 \label{HS-TC-2-CS}
\end{eqnarray}
The latter comes essentially from the T-pion channels $P_aP_b$ and
$SP_a$, while T-sigma one is suppressed by the relative velocity
squared, i.e. $(\sigma v_B)_{\rm ann}^{\rm SS}\sim v_B^2$.

Based upon a QCD analogy the T-strong Yukawa interactions are much
more intensive than the EW interactions, i.e. $g_{\rm TC}^N\gtrsim
1$, $g_{\rm TC}^N\gg g_{1,2},g_2^{\rm V}$ leading to a strong
dominance of pure TC (T-pion induced) annihilation channels in both
models I and II, such that
\begin{eqnarray}
 (\sigma v_B)_{\rm ann} \simeq (\sigma v_B)_{\rm ann}^{\rm TC}\,,
 \label{HS-TC-tot}
\end{eqnarray}
which makes the considering T-baryon DM model specific compared to
other standard SUSY-based DM models where $(\sigma v_B)_{\rm
ann}\sim \alpha_{\rm W}^2/M_{\chi}^2$, $\alpha_{\rm W}\simeq 1/30$
given by weak interactions only. Thus, more intense T-baryons
annihilation with extremely weak interactions with ordinary matter
(see below) makes them promising DM candidates alternative to
standard WIMPs. Note that the value (\ref{HS-TC-2-CS}) behaves as
forth power of $g_{\rm TC}^N$ leading to a large sensitivity of the
T-baryon mass scale extracted from the DM relic abundance data to
this parameter (see below).

\subsection{Annihilation of vector-like T-baryons: the low-symmetry phase}

\begin{figure}[h!]
 \centerline{\includegraphics[width=0.9\textwidth]{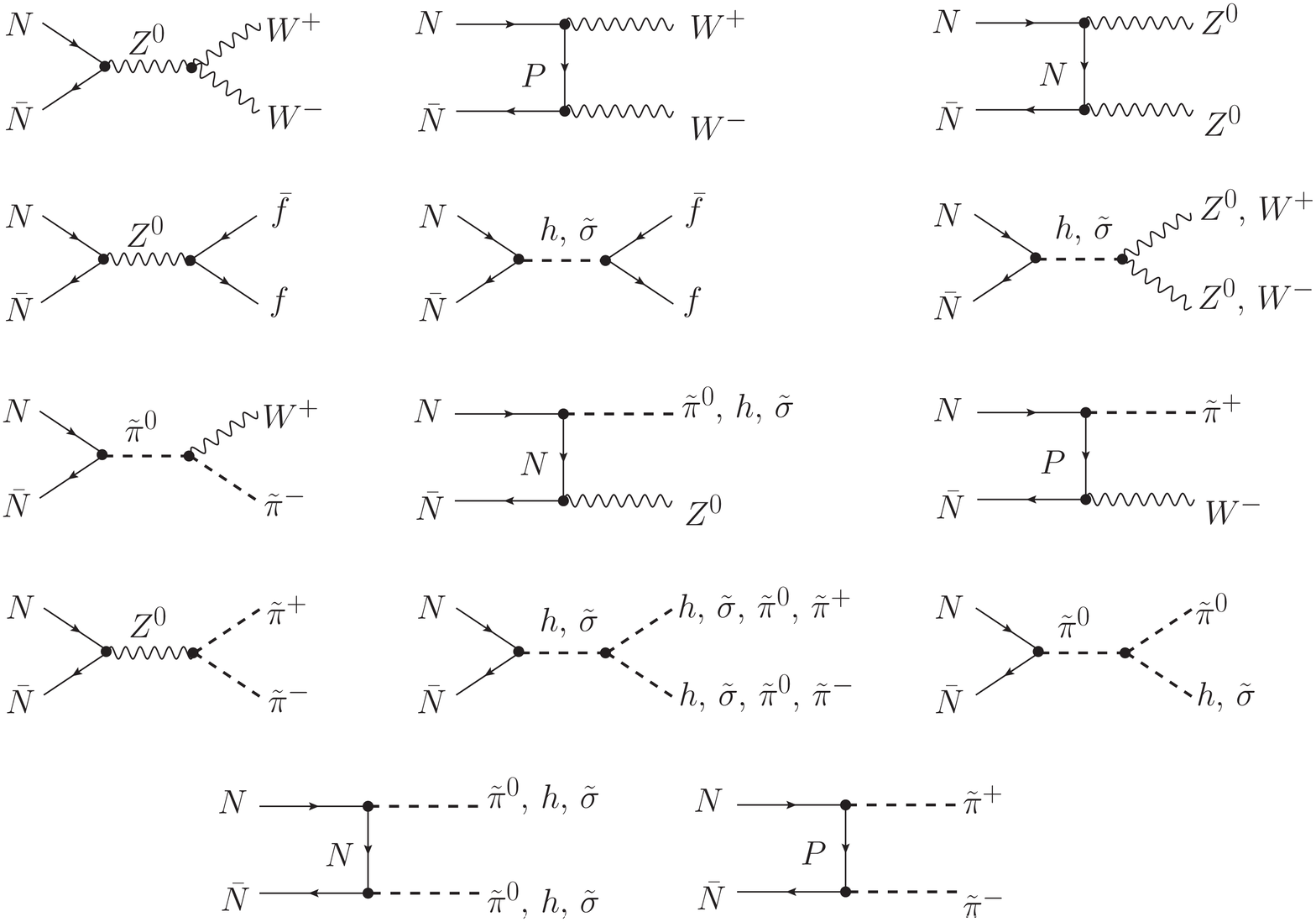}}
   \caption{\small Typical diagrams contributing to the T-neutron
   DM annihilation in the low-symmetry phase of the cosmological
   plasma $T<T_{\rm EW}$ including much later stages of structure
   formation and the present epoch. The model I
   (\ref{scenarios}) is implied.}
   \label{fig:DM-ann-halo}
\end{figure}

Again, consider first the phenomenologically appealing model I
(\ref{scenarios}) in detail. At lower temperatures, $T < T_{\rm EW}$
often referred to as to {\it the low-symmetry (LS) phase of the
cosmological plasma} the EW symmetry is broken and all the SM
fermions and gauge bosons acquire non-zeroth masses due to
non-trivial Higgs vev $v\simeq 246$ GeV. The latter scenario is
conventional for low-scale SUSY-based DM models with e.g. neutralino
LSP \cite{Belanger:2013oya}. This type of DM annihilation dynamics
would happen entirely after the EW phase transition epoch for rather
low mass T-baryons, i.e. $M_{B_{\rm T}}<3\,T_{\rm EW}\sim 600 \;
{\rm GeV}$ giving rise to cosmological consequences specific to the
considered vector-like TC model. The vector boson masses $m_{W,Z}$,
top quark mass $m_t$ together with $m_{\tilde \pi}$ and $M_{\tilde
\sigma}$ cannot be considered as negligible compared to the T-baryon
mass scale $M_{B_{\rm T}}$ any longer and have to be included making
respective calculations more involved than the ones in the HS phase.
Corresponding contributions to the kinetic annihilation cross
section $(\sigma v_B)_{\rm ann}$ in this case are listed in
Fig.~\ref{fig:DM-ann-halo}.

Here, only T-neutrons participate in the annihilation processes.
Indeed, T-protons have very small mean life-time (\ref{P-life}) so
they rapidly decay soon after the EW phase transition epoch and can
not substantially contribute to the T-baryon annihilation processes,
and thus to DM relic abundance formation in the low-symmetry phase.
The annihilation reactions shown in Fig.~\ref{fig:DM-ann-halo} also
happen at later stages of the Universe evolution during the
structure formation including the present epoch and thus are
relevant for ongoing indirect DM detection measurements.

In distinction to the HS phase, there are no pure EW annihilation
channels of the T-neutrons since vector boson channels may go also
through intermediate $h$ and $\tilde \sigma$ exchanges involving
T-strong Yukawa couplings and a small ${\tilde \sigma}h$ mixing
angle $\theta$. A straightforward calculation, however, reveals that
$s$-channel Higgs boson and T-sigma contributions in EW annihilation
channels are of the $v_B^2$ order including interference terms and
can be neglected.

The EW channels with SM fermion and gauge boson final states shown
in the first two lines in Fig.~\ref{fig:DM-ann-halo} are
\begin{eqnarray}
 && N\bar{N}\; \to\; {Z^0}^*,h^*,{\tilde \sigma}^*\; \to\; l{\bar l},\; q{\bar q}\,,
 \qquad N\bar{N}\; \to\; W^+W^-,\;Z^0Z^0 \,,
 \label{LS-EW}
\end{eqnarray}
respectively. The annihilation cross section into fermions is
\begin{eqnarray} \nonumber
 (\sigma v_B)_{\rm ann}^{f{\bar f}} &\simeq&
 \frac{1}{192\pi M_{B_{\rm T}} (4M_{B_{\rm T}}^2-m_Z^2)^2}
 \Big\{2M_{B_{\rm T}}^3(103g_1^4+6g_1^2g_2^2+63g_2^4)+ \\
 && (17g_1^4-6g_1^2g_2^2+9g_2^4)
 (2M_{B_{\rm T}}^2+m_t^2)\sqrt{M_{B_{\rm T}}^2-m_t^2}\Big\}\,,
 \label{sLS-EW-ff}
\end{eqnarray}
into $W^+W^-$ pair
\begin{eqnarray} \nonumber
 (\sigma v_B)_{\rm ann}^{W^+W^-} &\simeq& \frac{g_2^4}{64\pi M_{B_{\rm
 T}} m_W^4}\,\frac{(M_{B_{\rm T}}^2-m_W^2)^{3/2}}{(2M_{B_{\rm T}}^2-m_W^2)^2
 (4M_{B_{\rm T}}^2-m_Z^2)^2}\times \\
 && \Big\{ 4 M_{B_{\rm T}}^4 (12 m_W^4 - 4 m_W^2 m_Z^2 + m_Z^4) + \nonumber \\
 && 4 M_{B_{\rm T}}^2 m_W^2 (20 m_W^4 - 24 m_W^2 m_Z^2 + 5 m_Z^4) + \nonumber \\
 && m_W^4 (12 m_W^4 - 12 m_W^2 m_Z^2 + 5 m_Z^4) \Big\}\,,
 \label{sLS-EW-ww}
\end{eqnarray}
and into $Z^0Z^0$ pair
\begin{eqnarray}
 (\sigma v_B)_{\rm ann}^{Z^0Z^0} \simeq \frac{g_2^4(M_{B_{\rm T}}^2-m_Z^2)^{3/2}}{64\pi c_W^4 M_{B_{\rm
 T}} (2M_{B_{\rm T}}^2-m_Z^2)^2}
 \label{sLS-EW-zz}
\end{eqnarray}
Above we have neglected all the fermion masses assuming for
simplicity $m_{l,q}\ll M_{B_{\rm T}}$ except for top-quark mass
$m_t\simeq 173$ GeV. Also, vector boson and (pseudo)scalar masses
cannot be neglected and are kept here. The $N\bar{N}\to h^*,{\tilde
\sigma}^*\to l{\bar l},\; q{\bar q},\;W^+W^-$ processes and their
interference with pure EW channels have an order of $\sim v_B^2$. So
the $\tilde \sigma$ and $h$-mediated diagrams were neglected in the
integrated cross sections (\ref{sLS-EW-ff}), (\ref{sLS-EW-ww}) and
(\ref{sLS-EW-zz}).

The channels with mixed (gauge and (pseudo)scalar) final states
(third line in Fig.~\ref{fig:DM-ann-halo}) are
\begin{eqnarray}
 N\bar{N}\; \to \; W^{\pm}{\tilde \pi}^{\mp}\,, \qquad
 N\bar{N}\; \to \; Z^0{\tilde \pi}^0 \,, \qquad
 N\bar{N}\; \to \; Z^0{\tilde \sigma} \,, \qquad
 N\bar{N}\; \to \; Z^0 h \,,
 \label{LS-TC-1}
\end{eqnarray}
leading to the following contributions
\begin{eqnarray}
 (\sigma v_B)_{\rm ann}^{W^{\pm}{\tilde \pi}^{\mp}} &\simeq& \frac{g_2^2 (g_{\rm TC}^N)^2
 \Big[16M_{B_{\rm T}}^4-8M_{B_{\rm T}}^2(m_{\tilde \pi}^2+m_W^2)+
 (m_{\tilde \pi}^2-m_W^2)^2\Big]^{3/2}}{32\pi M_{B_{\rm
 T}}^4(4M_{B_{\rm T}}^2-m_{\tilde \pi}^2)^2(4M_{B_{\rm T}}^2-m_{\tilde
 \pi}^2-m_W^2)^2}\times \label{LS-TC-1-CS-Wpi} \\
 && \Big\{(4M_{B_{\rm T}}^2-m_{\tilde \pi}^2)^2+2M_{B_{\rm T}}^2m_W^2\Big\}\,, \nonumber \\
 (\sigma v_B)_{\rm ann}^{Z^0{\tilde \pi}^0} &\simeq& \frac{g_2^2 (g_{\rm TC}^N)^2
 \Big[16M_{B_{\rm T}}^4-8M_{B_{\rm T}}^2(m_{\tilde \pi}^2+m_Z^2)+
 (m_{\tilde \pi}^2-m_Z^2)^2\Big]^{3/2}}{128\pi c_W^2 M_{B_{\rm
 T}}^4(4M_{B_{\rm T}}^2-m_{\tilde \pi}^2-m_Z^2)^2}\,,  \label{LS-TC-1-CS-Zpi}
\end{eqnarray}
\begin{eqnarray}
 (\sigma v_B)_{\rm ann}^{Z^0{\tilde \sigma}} &\simeq& \frac{g_2^2 (g_{\rm
 TC}^N)^2\,c_\theta^2\,
 \sqrt{16M_{B_{\rm T}}^4-8M_{B_{\rm T}}^2(M_{\tilde \sigma}^2+m_Z^2)+
 (M_{\tilde \sigma}^2-m_Z^2)^2}}{128\pi c_W^2 M_{B_{\rm
 T}}^4(4M_{B_{\rm T}}^2-M_{\tilde \sigma}^2-m_Z^2)^2}\times \nonumber \\
 && \Big\{ 16M_{B_{\rm T}}^4-8M_{B_{\rm T}}^2(M_{\tilde \sigma}^2-2m_Z^2)+
 (M_{\tilde \sigma}^2-m_Z^2)^2\Big\}\,, \label{LS-TC-1-CS-Zsig} \\
 (\sigma v_B)_{\rm ann}^{Z^0h} &\simeq& \frac{g_2^2 (g_{\rm
 TC}^N)^2\,s_\theta^2\,
 \sqrt{16M_{B_{\rm T}}^4-8M_{B_{\rm T}}^2(m_h^2+m_Z^2)+
 (m_h^2-m_Z^2)^2}}{128\pi c_W^2 M_{B_{\rm
 T}}^4(4M_{B_{\rm T}}^2-m_h^2-m_Z^2)^2}\times \nonumber \\
 && \Big\{ 16M_{B_{\rm T}}^4-8M_{B_{\rm T}}^2(m_h^2-2m_Z^2)+
 (m_h^2-m_Z^2)^2\Big\}\,, \label{LS-TC-1-CS-ZH}
\end{eqnarray}
where $m_h\simeq 126$ GeV is the Higgs boson mass \cite{Higgs}.

Finally, last two lines in Fig.~\ref{fig:DM-ann-halo} represent
diagrams with T-strong final states (T-pions and T-sigma) as well as
the Higgs boson:
\begin{eqnarray} \nonumber
 && N\bar{N}\; \to\; {\tilde \pi}^+{\tilde \pi}^-\,, \qquad
 N\bar{N}\; \to\; {\tilde \pi}^0{\tilde \pi}^0 \,, \qquad
 N\bar{N}\; \to\; {\tilde \pi}^0{\tilde \sigma} \,, \qquad
 N\bar{N}\; \to\; {\tilde \pi}^0 h \,, \\
 && \qquad\qquad\qquad N\bar{N}\; \to\; h{\tilde \sigma} \,, \qquad
 N\bar{N}\; \to\; {\tilde \sigma}{\tilde \sigma} \,, \qquad
 N\bar{N}\; \to\; hh \,.
 \label{LS-TC-2}
\end{eqnarray}
The relevant contributions to the total cross section are
\begin{eqnarray}
 (\sigma v_B)_{\rm ann}^{{\tilde \pi}^+{\tilde \pi}^-} &\simeq&
 \frac{(M_{B_{\rm T}}^2-m_{\tilde \pi}^2)^{3/2}}{16\pi M_{B_{\rm
 T}}}\left(\frac{g_2^4}{(4M_{B_{\rm T}}^2-m_Z^2)^2}+
 \frac{4(g_{\rm TC}^N)^4}{(2M_{B_{\rm T}}^2-m_{\tilde
 \pi}^2)^2}\right)\,,
 \label{LS-TC-2-CS-pipi} \\
 (\sigma v_B)_{\rm ann}^{{\tilde \pi}^0h} &\simeq&
 \frac{(g_{\rm TC}^N)^2\,s_\theta^2\,
 \sqrt{16M_{B_{\rm T}}^4-8M_{B_{\rm T}}^2(m_h^2+m_{\tilde \pi}^2)+
 (m_h^2-m_{\tilde \pi}^2)^2}}{256\pi M_{B_{\rm T}}^4 M_Q^2
 (4M_{B_{\rm T}}^2-m_{\tilde \pi}^2)^2
 (4M_{B_{\rm T}}^2-m_{\tilde \pi}^2-m_h^2)^2} \times \nonumber \\
 && \Big( g_{\rm TC}^Q M_{B_{\rm T}}
 (m_{\tilde \pi}^2-m_h^2)(4M_{B_{\rm T}}^2-m_{\tilde \pi}^2-m_h^2)
 + \nonumber \\
 && 2 g_{\rm TC}^N M_Q (4M_{B_{\rm T}}^2-m_{\tilde \pi}^2)
 (4M_{B_{\rm T}}^2+m_{\tilde \pi}^2-m_h^2)\Big)^2\,,
 \label{LS-TC-2-CS-pih} \\
 (\sigma v_B)_{\rm ann}^{{\tilde \pi}^0{\tilde \sigma}} &\simeq&
 \frac{(g_{\rm TC}^N)^2\,c_\theta^2\,
 \sqrt{16M_{B_{\rm T}}^4-8M_{B_{\rm T}}^2(M_{\tilde \sigma}^2+m_{\tilde \pi}^2)+
 (M_{\tilde \sigma}^2-m_{\tilde \pi}^2)^2}}{256\pi M_{B_{\rm T}}^4 M_Q^2
 (4M_{B_{\rm T}}^2-m_{\tilde \pi}^2)^2 (4M_{B_{\rm T}}^2-m_{\tilde \pi}^2-
 M_{\tilde \sigma}^2)^2} \times \nonumber \\
 && \Big( g_{\rm TC}^Q M_{B_{\rm T}}
 (m_{\tilde \pi}^2-M_{\tilde \sigma}^2)
 (4M_{B_{\rm T}}^2-m_{\tilde \pi}^2-M_{\tilde \sigma}^2) + \nonumber \\
 && 2 g_{\rm TC}^N M_Q (4M_{B_{\rm T}}^2-m_{\tilde \pi}^2)
 (4M_{B_{\rm T}}^2+m_{\tilde \pi}^2-M_{\tilde \sigma}^2)\Big)^2\,,
 \label{LS-TC-2-CS-pisig}
\end{eqnarray}
while scalar (${\tilde \sigma}{\tilde \sigma}$, $hh$ and ${\tilde
\sigma}{\tilde \sigma}$) and pseudoscalar ${\tilde \pi}^0{\tilde
\pi}^0$ channels are suppressed as $\sim v_B^2$ due to a cancelation
between $t$- and $u$-channel diagrams similar to the HS phase where
the T-pion and mixed T-pion/T-sigma channels dominate in the total
cross section. In order to turn to model II with broken $SU(2)_{\rm
V}\not=SU(2)_{\rm W}$ symmetry, one has to make the following
replacements $m_Z\to m_{Z'},\,m_W\to m_{W'},\,g_2\to g_2^{\rm V}$ in
the above formulae (\ref{sLS-EW-ff})--(\ref{LS-TC-2-CS-pisig}).

In the above expressions, $M_{B_{\rm T}}\simeq 3M_Q$, $m_{\tilde
\pi}$, $M_{\tilde \sigma}$, and $g_{\rm TC}^{N,Q}$ are kept as free
parameters to be constrained from (collider and astrophysics)
phenomenology. Due to a rather strong inequality $g_{\rm
TC}^{N,Q}\gg g_{1,2},g_2^{\rm V}$ characteristic for the new
strongly-coupled dynamics under discussion, the ${\tilde
\pi}^+{\tilde \pi}^-$-channel $(\sigma v_B)_{\rm ann}^{{\tilde
\pi}^+{\tilde \pi}^-}$ and mixed scalar-pseudoscalar channels
$(\sigma v_B)_{\rm ann}^{{\tilde \pi}^0h}$ and $(\sigma v_B)_{\rm
ann}^{{\tilde \pi}^0{\tilde \sigma}}$ dominate the total T-baryon
annihilation cross section in the LS phase for not very large
$M_{B_{\rm T}}\lesssim 600$ GeV. Note that all the annihilation
cross section in the HS and LS phases behave as $(\sigma v_B)_{\rm
ann}\sim M_{B_{\rm T}}^{-2}$ in the limit of large $M_{B_{\rm T}}$.

It is worth to mention here that in practice depending on T-baryon
mass $M_{B_{\rm T}}$ and intensity of annihilation processes an
intermediate scenario when the annihilation epoch starts in the HS
phase and terminates in the LS phase is possible. In this case, the
residual T-baryon density may be strongly correlated with details of
EW phase transition epoch, as well as on relative annihilation
intensities and time periods during which the plasma ``lives'' in
each of the phases. Such an analysis is certainly much more
complicated and invokes a larger amount of unknowns into the
problem. However, if the considering vector-like TC model is
phenomenologically justified, a more detailed analysis of
cosmological evolution of the T-baryon component along these lines
would be necessary. Instead, we are focused here on two simplistic
toy-scenarios capturing basic dynamics of vector-like T-baryons in
cosmological environment when their annihilation happens either in
the HS or LS phase entirely.

\section{Cosmological evolution of vector-like T-baryons}

The T-proton lifetime (\ref{P-life}) has an order of Universe age at
the EW phase transition epoch $t_{\rm EW}$. So after this epoch the
T-proton remnants quickly decay into T-neutrons and light SM
fermions without leaving significant traces on cosmological
evolution of cold DM and ordinary matter. Indeed, due to a rather
small T-nucleon mass splitting (\ref{deltaM-num}) the amount of
T-baryons frozen off the cosmological plasma will be approximately
equal to the T-neutron/anti-T-neutron DM density extrapolated back
to the freeze-out epoch. So T-proton/anti-T-proton decay processes
and relative fraction of T-protons and T-neutrons at freeze-out time
scale can be disregarded in analysis of the DM relic abundance, at
least, to the first approximation. Let us consider two possible
scenarios of T-baryon DM relic abundance formation -- symmetric and
asymmetric DM cases.

\subsection{Scenario I: Thermal freeze-out of symmetric T-baryon Dark Matter}

Consider thermal evolution of T-baryon density in the cosmological
plasma in the case of symmetric DM, i.e. when number densities of
T-baryons and anti-T-baryons are (at least, approximately) equal to
each other at all stages of Universe evolution until the present
epoch, i.e.
\begin{eqnarray}
n_{B_{\rm T}}\simeq n_{\bar{B}_{\rm T}}\,, \qquad n_{B_{\rm
T}}-n_{\bar{B}_{\rm T}}\ll n_{B_{\rm T}}\,,
\end{eqnarray}
so the chemical potential of T-baryon plasma can be neglected to a
sufficiently good accuracy. In the considering model the T-baryons
are Dirac particles so $B_{\rm T}$ do not coincide with
$\bar{B}_{\rm T}$, while these particles are always produced and
annihilate in $B_{\rm T}\bar{B}_{\rm T}$ pairs, at least, at
relevant temperatures $T<M_{B_{\rm T}}$.

For details of the thermal evolution and freeze out of symmetric
heavy relic, see e.g. Ref.~\cite{Steigman:2012nb} and references
therein, while here we repeat just a few relevant formulas. The
irreversible T-baryon annihilation process in the cosmological
plasma is initiated at the moment $t_0$ and temperature
$T_0=T(t_0)$, when the mean energy of relativistic quarks and
leptons is comparable with T-baryon mass scale $M_{B_{\rm T}}$ in
the radiation-dominated epoch, i.e. $\bar \varepsilon_f \simeq
3\,T_0 \simeq M_{B_{\rm T}}$ and the Hubble parameter is $H=1/2t$.
The chemical equilibrium with respect to T-baryon
annihilation/production processes breaks down at this moment, and
the residual density of T-baryons in the plasma $n_{B_{\rm
T}}=n_{B_{\rm T}}(t)$ at later times $t\gg t_0$ is then described by
a standard solution of the Boltzmann evolution equations
\cite{Steigman:2012nb}.

In the HS phase of the Universe evolution (model I), the T-baryon
annihilation epoch effectively begins at the moment of physical time
\begin{eqnarray}
 t_0\simeq
 \frac{1}{4T_{0}^2}\left(\frac{3}{2\pi Gw}\right)^{1/2}\ll t_{\rm
 EW}\,, \qquad T_0\simeq \frac{M_{B_{\rm T}}}{3}\,,
 \label{tann}
\end{eqnarray}
where $G$ is the gravitational constant, $w=g_*(T)\pi^2/30$ is the
statistical weight of cosmological plasma at the T-baryon
annihilation epoch, and $g_*=g_*(T)$ is the effective number of
relativistic d.o.f. in the plasma. At high temperatures $T>T_{\rm
EW}\sim 200$ GeV before the EW phase transition, in the SM with one
Higgs doublet we have $g_*=106.75$. At lower temperatures, $T <
m_W$, the value $g_* = 86.25$ can be used. The steepest change in
$g_*$ happens around the QCD phase transition $T_{\rm QCD}\sim 100$
MeV when it drops down to $g_*\simeq 10$, but the latter does not
affect the heavy T-baryons evolution at late times since they have
already decoupled from the plasma while annihilation rate is
practically negligible on average.

Assuming that the annihilation epoch occurs entirely in the HS
phase, it should terminate before the EW phase transition time as
soon as T-baryons drop off the chemical equilibrium. Then the freeze
out of heavy T-baryons happens at $t=t_1$ when temperature of the
Universe $T_1=T_1(t_1)\gtrsim T_{\rm EW}$ given by the standard
formula
\begin{eqnarray}
T_1\simeq \frac{M_{B_{\rm T}}}{\log\Big(\frac{g_{B_{\rm T}}M_{B_{\rm
T}}M_{\rm PL}^*(\sigma v_B)_{\rm ann}}{(2\pi)^{3/2}}\Big)}\ll
M_{B_{\rm T}}\,, \qquad M_{B_{\rm T}}\ll M_{\rm PL}^* \,,
\label{freeze-end}
\end{eqnarray}
valid to a logarithmic accuracy. Here, $M_{\rm PL}^*=M_{\rm
PL}/1.66\sqrt{g_*}$ is the reduced Planck mass, and $g_{B_{\rm T}}$
is the number of T-baryon d.o.f. The phase of cosmological plasma
where the DM gets effectively frozen out, i.e. an actual relation
between $T_1$ and the EW phase transition temperature $T_{\rm EW}$,
depends on details of a DM scenario, or on typical $M_{B_{\rm T}}$
and $(\sigma v_B)_{\rm ann}$ values in our case. Typical weak
interactions strength leads to a crude order-of-magnitude estimate
$(\sigma v_B)_{\rm ann}\sim \alpha_{\rm W}^2/M_{\rm EW}^2$,
$\alpha_{\rm W}\simeq 1/30$ , such that $T_1\simeq M_{B_{\rm
T}}/20$. Additional T-strong annihilation channels may affect this
estimate but only logarithmically in respective cross section (or
mass).

Under the basic assumption that the DM in the present epoch
$t=t_{\rm U}$ consists mostly of heavy particles of one type, e.g.
T-neutrons, the condition on current mass density of the DM
\begin{eqnarray}
 \rho_{B_{\rm T}}(t_{\rm U})=M_{B_{\rm
T}}n_{B_{\rm T}}(t_{\rm U})\simeq \rho_{\rm DM}(t_{\rm U})\,,
 \label{ysl}
\end{eqnarray}
provides the canonical constraint on thermally averaged kinetic
annihilation cross section known from Ref.~\cite{Steigman:2012nb}
\begin{eqnarray}
 (\sigma v_B)_{\rm ann}^{\rm DM}\simeq 2.0\times 10^{-9}\; {\rm
 GeV}^{-2}\,.
 \label{sigvann}
\end{eqnarray}
In terms of the latter, the relic DM abundance is
\begin{eqnarray}
\Omega_{\rm DM}\simeq 0.2\Big[(\sigma v_B)_{\rm ann}^{\rm
DM}/(\sigma v_B)_{\rm ann}^{\rm th}\Big]\,.
\end{eqnarray}
This formula can be used for determination of the T-baryon mass
scale $M_{B_{\rm T}}$ as long as a theoretical prediction for the
annihilation cross section $(\sigma v_B)_{\rm ann}^{\rm th}$ is
given.

By comparing the above astrophysical constraint with the theoretical
cross section given by Eq.~(\ref{HS-TC-2-CS}) one extracts the lower
bound for the T-baryon mass scale
\begin{eqnarray}
 M_{B_{\rm T}} \gtrsim 5 \; {\rm TeV}\,, \qquad g_{\rm TC}^N\gtrsim 1.0
 \label{M-TB}
\end{eqnarray}
Clearly, this bound is consistent with naive QCD scaling hypothesis
(\ref{QCD-an}). An actual T-baryon mass estimate may vary in a very
broad range from a few TeV to a few tens of TeV due to a strong
dependence of the cross section on T-strong Yukawa coupling,
$(\sigma v_B)_{\rm ann}\sim (g_{\rm TC}^N)^4$. Then typically the
irreversible T-baryon annihilation starts at the temperature
$T_0\gtrsim 1-10\;{\rm TeV}$ and terminates at $T_1\gtrsim
0.1-1\;{\rm TeV}$ or higher. So indeed the annihilation epoch occurs
entirely in the HS-phase for the model I (\ref{scenarios})
demonstrating consistency of the considering scenario of T-baryon
relic formation. For the model II, a high-scale $SU(2)_{\rm V}$
symmetry is likely to be broken at temperatures $T<M_{B_{\rm T}}/3$
so in this case the annihilation occurs in the low-symmetry (w.r.t.
$SU(2)_{\rm V}$) phase. Note, somewhat lower $M_{B_{\rm T}}$
estimates are also possible for smaller $g_{\rm TC}^N$ values in
both models I and II, but then mixed EW+TC and pure EW channels
become important, such that the mean value $M_{B_{\rm T}}$ never
goes below 3 TeV corresponding to ``switched off'' Yukawa TC
interactions, i.e. $g_{\rm TC}^N\to 0$.

Note, the estimate (\ref{M-TB}) should be taken with care since they
have been obtained under a few generic conditions which can be
summarized as follows:
\begin{itemize}
 \item the DM is made entirely of heavy T-neutrons (i.e. contributions
 to the DM mass density from possible lighter components is negligibly
 small);
 \item the T-baryon number is conserved similarly to the baryon
 number;
 \item the T-protons decay very fast soon after the EW phase
 transition epoch, and their decays do not affect the cosmological
 evolution of the neutrino gas and CMB;
 \item the neutrino gas evolution is adiabatic;
\end{itemize}
and a few model-specific assumptions:
\begin{itemize}
 \item there is a significant splitting between the chiral symmetry
 breaking scale and the EW symmetry breaking scale allowing for a
 strong hierarchy between $M_{B_{\rm T}}$ and $M_{\rm EW}$;
 \item the number densities of T-baryons and anti-T-baryons are the
 same to a good approximation;
 \item the T-baryon distribution is roughly homogeneous so
 a possible extra loss of the DM due to its annihilation
 in dense regions of the Universe during structure formation
 epoch is neglected.
\end{itemize}
Clearly, a more involved analysis of relic T-baryon abundance
evolution lifting out one or more of the specific assumptions would
be necessary.

\begin{figure*}[!h]
\begin{minipage}{0.49\textwidth}
 \centerline{\includegraphics[width=1.0\textwidth]{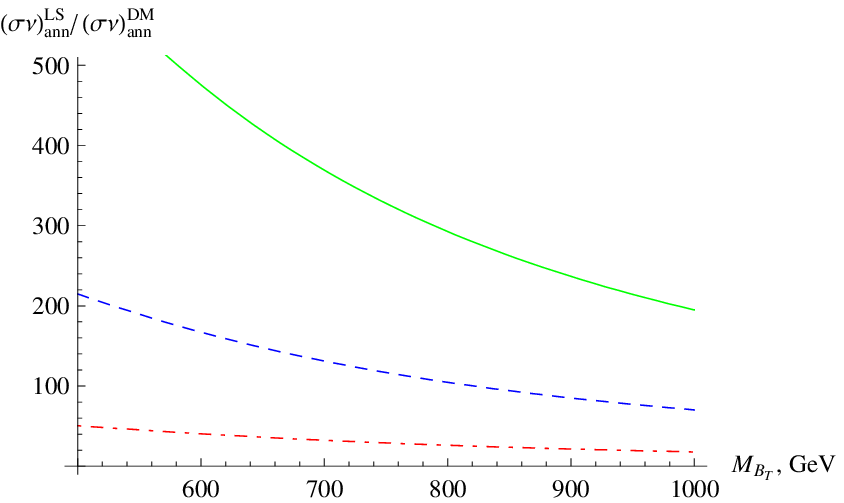}}
\end{minipage}
\begin{minipage}{0.49\textwidth}
 \centerline{\includegraphics[width=1.0\textwidth]{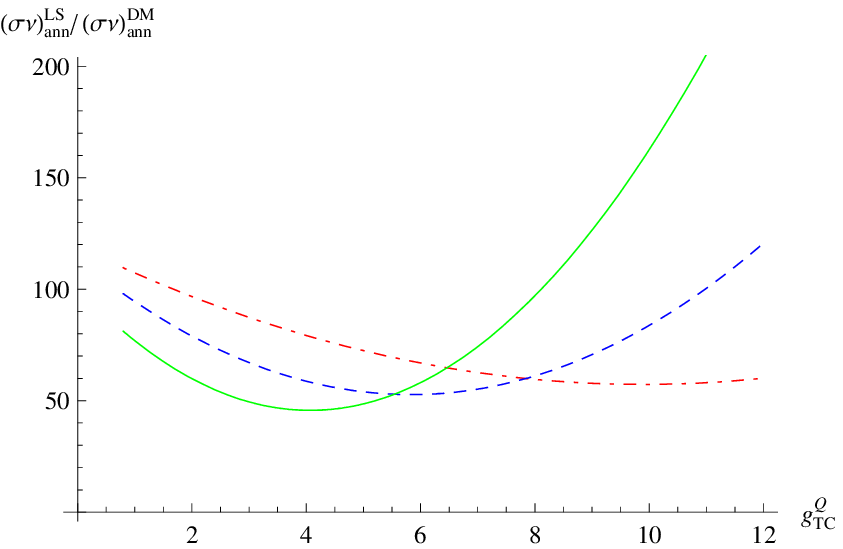}}
\end{minipage}
   \caption{
\small Dependence of the ratio of kinetic T-baryon cross section in
the LS phase for the model I $(\sigma v_B)_{\rm ann}^{\rm LS}$
(\ref{LS-CS}) to the kinetic cross section of the DM $(\sigma
v_B)_{\rm ann}^{\rm DM}\simeq 2.0\times 10^{-9}\; {\rm GeV}^{-2}$
required by observations on T-neutron mass $M_{B_{\rm T}}$ (left)
and on T-quark Yukawa coupling $g_{\rm TC}^{Q}$ (right). In both
panels we adopt the ``no ${\tilde \sigma}h$-mixing'' limit
$s_\theta=0,\, M_{\tilde \sigma}=\sqrt{3}m_{\tilde \pi}$ for
simplicity. In the left panel, three different $g_{\rm
TC}^N=1.0,1.4,1.8$ values correspond to dash-dotted, dashed and
solid lines, respectively, and fixed $m_{\tilde \pi}=250$ GeV and
$g_{\rm TC}^Q=2.0$. In the right panel, $m_{\tilde \pi}=150,200,250$
GeV correspond to dash-dotted, dashed and solid lines, respectively,
and fixed $M_{B_{\rm T}}=400$ GeV and $g_{\rm TC}^N=1.0$.}
 \label{fig:fig-ratio}
\end{figure*}

Finally, consider formation of the relic symmetric DM density in the
LS phase of the cosmological plasma for the model I set-up. It
appears to be very hard to realize such a scenario with $M_{B_{\rm
T}}\lesssim 600$ GeV due to a large $(\sigma v_B)_{\rm ann}^{{\tilde
\pi}^+{\tilde \pi}^-}$ contribution for $g_{\rm TC}^N\gtrsim 1$ and
$M_{B_{\rm T}}\gg m_{\tilde \pi}$ (see also
Fig.~\ref{fig:fig-ratio}). Formally, it may still be possible to
tune $m_{\tilde \pi}$ and $g_{\rm TC}^N$ in a special way by
minimizing the total LS cross section given by
 \begin{eqnarray}
(\sigma v_B)_{\rm ann}^{\rm LS} \simeq (\sigma v_B)_{\rm
ann}^{{\tilde \pi}^+{\tilde \pi}^-} + (\sigma v_B)_{\rm
ann}^{{\tilde \pi}^0{\tilde \sigma}} + (\sigma v_B)_{\rm
ann}^{{\tilde \pi}^0h}\,,
 \label{LS-CS}
 \end{eqnarray}
but the corresponding TC model would be unnatural. In the model II,
there is a larger freedom in relative (rather fine) tuning between
large $m_{W'},\,m_{Z'}$ and $M_{B_{\rm T}}$ parameters capable of
reducing the annihilation cross section. This, however, cannot bring
$M_{B_{\rm T}}$ down significantly without spoiling EW precision
constraints and keeps it roughly at the same level (\ref{M-TB}).
Other possibility would be to have the annihilation epoch in the
mixed HS/LS phase by relaxing the constraint on the T-baryon mass
scale $M_{B_{\rm T}}\lesssim 600$ GeV. The LS and HS annihilation
cross sections become comparable for $M_{B_{\rm T}}\gtrsim 3$ TeV,
such that in this case most of the T-baryon annihilation epoch
happens in the HS phase again justifying the above estimate
(\ref{M-TB}).

Thus, we come to a conclusion that the most likely scenario with
symmetric T-baryon DM can only be realized for the T-baryon mass
scale of at least a few TeV or more, so the corresponding DM
particles should be very heavy.

\subsection{Scenario II: Asymmetric vector-like T-baryon Dark Matter}

As was argued above the LS annihilation cross section (\ref{LS-CS})
(model I) becomes too high for low mass T-baryons. Indeed, in
Fig.~\ref{fig:fig-ratio} we show typical parameter dependencies of
the ratio $(\sigma v_B)_{\rm ann}^{\rm LS}/(\sigma v_B)_{\rm
ann}^{\rm DM}$. It turns out that the LS cross section is by far
larger than is required by observations at least by a factor of
$10-100$ or more. This means that relatively light T-neutrons and
anti-T-neutrons $M_{B_{\rm T}}\lesssim 600$ GeV most probably have
quickly annihilated off in the cosmological plasma by the time of
their freeze out shortly after the EW phase transition epoch. In
this case, in order to provide the observable DM abundance it is
natural to assume the existence of {\it a T-baryon asymmetry} in a
complete analogy with the typical baryon asymmetry. So, the bulk of
observable DM density is essentially given in terms of the T-baryon
asymmetry $\Delta n_{B_{\rm T}}$, i.e.
\begin{eqnarray}
\rho_{\rm DM}(t_{\rm U})\simeq \Delta n_{B_{\rm T}} M_{B_{\rm T}}\,,
\qquad \Delta n_{B_{\rm T}} \equiv n_{B_{\rm T}}-n_{\bar{B}_{\rm
T}}\,,
\end{eqnarray}
even though it has been negligible in the beginning of the T-baryon
annihilation epoch, $\Delta n_{B_{\rm T}}\ll n_{B_{\rm T}}(t_0)$.
This is the so-called asymmetric DM (ADM) model which has been
previously studied in other TC/compositeness scenarios (see e.g.
Refs.~\cite{Masina:2012hg,Lewis:2011zb,DelNobile:2011je}), and it
can be realized in the considering vector-like TC model as well.

Introducing a fractional asymmetry of T-neutrons $N$ and
anti-T-neutrons $\bar N$ in the cosmological plasma as
\begin{eqnarray}
r\equiv \frac{n(\bar{N})}{n(N)}\,,\qquad 0<r<1\,,
\end{eqnarray}
one could estimate its time evolution via the detailed analysis of
the system of coupled Boltzmann equations performed in
Refs.~\cite{Graesser:2011wi,Iminniyaz:2011yp}. The late-time
fractional asymmetry $r_{\infty}$ for light T-neutrons turns out to
be exponentially suppressed in most of the vector-like TC parameter
space, i.e.
\begin{eqnarray}
r_{\infty}\sim \exp\Big\{-2\frac{(\sigma v_B)_{\rm ann}^{\rm
LS}}{(\sigma v_B)_{\rm ann}^{\rm DM}}\Big\}\ll 1
\end{eqnarray}
for the dominant $s$-wave Dirac T-neutron annihilation processes
considered above in the LS annihilation (light T-neutron, model I).
Having typically large ratios of the cross sections illustrated in
Fig.~\ref{fig:fig-ratio} one concludes that there practically no
light anti-T-neutrons remain in modern Universe similar to ordinary
antibaryons. The T-neutrons by themselves cannot produce any
annihilation-like signal expected to be constrained by indirect
detection measurements while they may have a certain impact on
direct measurements \cite{Goodman:1984dc}.

Similarly to the HS phase annihilation, in the vector-like T-neutron
ADM scenario the T-baryon mass scale is expected to be above the EW
scale $M_{B_{\rm T}}\gtrsim 200$ GeV such that the chemical
decoupling of the ADM occurs when DM is non-relativistic while SM
fermions are still relativistic. Then one could assume a tight
relation between ordinary baryon and T-baryon asymmetries typically
considered in ADM scenarios which translates into a relation between
the number densities of the visible matter and T-neutron abundances.
In this case depending on details of the chemical equilibrium the
T-baryon mass scale is expected to be rather high $M_{B_{\rm
T}}\sim$ TeV as was advocated in Ref.~\cite{Buckley:2010ui} which is
consistent with the suggested vector-like ADM scenario having the
T-neutron annihilation epoch in the LS (or mixed HS+LS) phase.

For relatively low $g_{\rm TC}^N\sim 1$ and high T-neutron masses
$M_{B_{\rm T}}\gtrsim 1$ TeV the ratio of the cross sections goes
down
\begin{eqnarray}
1<\frac{(\sigma v_B)_{\rm ann}^{\rm LS}}{(\sigma v_B)_{\rm ann}^{\rm
DM}}\lesssim 10\,,
\end{eqnarray}
and can therefore accommodate the {\it partially ADM scenario}
$0<r_{\infty}<1$ with heavy T-neutrons allowing for many attractive
features. The same is true for the case of HS annihilation in the
model I and LS annihilation in the model II. In particular, having a
small but non-zeroth relic density of anti-T-neutrons would open up
immediate opportunities for indirect detection measurements of DM
annihilation products from galactic cores and compact stars. Also,
this scenario is a particular case of self-interacting DM model
which allows to avoid problematic cusp-like DM density profile in
the central regions of galactic haloes leading to a core-like DM
distribution favored by astrophysical observations
\cite{Wandelt:2000ad}. In the considering scenario, the intensive
annihilation together with elastic $NN$ scattering play an important
role of self-interactions of DM particles allowing for necessary
adjustments of the DM density profile. Indeed, in the early
structure formation epoch an overdensity of DM in cusp-like regions
has been eliminated by intensive annihilation processes such that
they do not exist today.

Another interesting astrophysical implication of the partially ADM
scenario in general is a possible thermalization of the cosmological
medium in the beginning of the structure formation epoch. Indeed,
the growth of structures is accompanied by a substantial increase in
DM density in the central regions of haloes. Having a large cross
section (\ref{LS-CS}) the intensity of $N{\bar N}$ annihilation
processes has gone up in that epoch and further reduced the amount
of anti-T-neutrons which have survived after the T-neutrons
freeze-out. The annihilation products could be capable of
thermalization of the medium at $z\sim 10$ or somewhat earlier which
may have serious observational consequences. Some small remnants of
anti-T-neutron density could have survived such a ``second T-baryon
annihilation epoch'' and remain today providing possible
observational signatures of their annihilation with T-neutrons.
Certainly, a thorough analysis involving simulations of the
structure formation epoch together with T-baryon annihilation and DM
formation details is required.

Finally, intensive vector-like T-neutron DM annihilation allows to
explain why the DM in the Galactic halo is much more tepid than
ordinary CDM models predict. Corresponding problem is typically
tagged as the missing satellite problem. Due to a much higher
temperature of the DM in the Galactic halo, the observed number of
dwarf galaxies is by an order of magnitude smaller while the DM
density in the halo cores is much smaller then CDM WIMP-based
simulations predict \cite{Mateo:1998wg,Moore:1999nt,Zavala:2012us}.
In the considering vector-like TC model in the case of partial ADM
one finds an interesting opportunity for such a ``tepid'' DM.
Indeed, at the initial stages of structure formation slower (colder)
DM particles in the central cusp-like regions have annihilated off
while faster particles moving in less dense regions at the Galactic
periphery according to the Jeans instability criterion could have
survived until today. So this proposal could be an efficient
alternative to warm DM models with low mass WIMPs ${\cal O}(1)$ GeV
aimed at resolving the above issues.

Thus, the vector-like ADM scenario with a relatively high
annihilation rate of heavy T-neutrons offers a few appealing
possibilities compared to traditional practically non-interacting
WIMP-miracle. The basic opportunities mentioned above inherent to
the considered T-neutron DM scenario should further be explored
quantitatively.

\section{Direct T-neutron detection constraints}

At last, let us consider one of the most important constraints on
the vector-like TC model with Dirac T-baryons -- the direct DM
detection limits. Among them the data on spin-independent (SI)
component of the elastic WIMP-nucleon cross section from CDMS II
\cite{Ahmed:2009zw} and, especially, XENON100 \cite{Aprile:2012nq}
experiments provide the most stringent model-independent exclusion
limits. Indeed, the elastic T-neutron-nucleon scattering goes via a
T-neutron vector coupling to the $Z$ boson defined in
Eq.~(\ref{L-QV}). This leads to a sizable SI scattering cross
section off nuclei which needs to be compared to the data.

Let us consider the Dirac T-neutron-nucleon scattering in the
non-relativistic limit $v_N\ll 1$. Previously, a similar process has
been investigated in the case of Dirac vector-like neutralino DM
(see e.g. Refs.~\cite{our-Higgsino,Chun:2009zx,Buckley:2013sca}), so
we do not go into details of explicit calculations here. Following
to Ref.~\cite{Buckley:2013sca} the SI spin-averaged
T-neutron-nucleus cross section reads
\begin{eqnarray}
\sigma_{\rm SI}=\frac{\mu^2}{16\pi M_{B_{\rm T}}^2
m_A^2}\left(\frac14 \sum_{\rm spins} |{\cal M}_{\rm SI}|^2\right)\,,
\end{eqnarray}
where $\mu$ is the reduced mass and $m_A$ is the mass of the target
nucleus. In the considering case at small momentum transfers $q^2\ll
m_Z^2$ the effective operator for T-neutron scattering off quarks
through $Z$-exchange is described by vector couplings only
\begin{equation}
{\cal O}^q_Z=\delta_Z\,\lambda_{qV}\,\frac{ig_2^2}{2c_W^2
m^2_Z}\,\bar{N}\gamma^{\mu}N\cdot \bar{q}\gamma_{\mu}q\,,
\end{equation}
where $q=u,d$ quarks in a nucleon, and the standard vector $Zq$
couplings are
\begin{equation}
\lambda_{qV}=\frac{g_2}{c_W}\,\big[t^3_{qL}-2Q_q s_W^2\big]\,.
\end{equation}
This leads to a squared matrix element (cf.
Ref.~\cite{Buckley:2013sca})
\begin{eqnarray}
&&\frac14 \sum_{\rm spins} |{\cal M}_{\rm SI}|^2 \simeq
16\,\frac{M_{B_{\rm T}}^2 m_A^2}{m_Z^4}\,\frac{g_2^4}{4c_W^4}\;
\delta_Z^2 \, \left[\,\sum_{q=u,d}\lambda_{qV}[Z B_{qV}^p + (A-Z)
B_{qV}^n]\right]^2\,,
\end{eqnarray}
where $B_{uV}^p=B_{dV}^n=2$ and $B_{uV}^n=B_{dV}^p=1$ are $u,d$
quark multiplicities in a proton $p$ and neutron $n$, and $Z$ and
$A$ correspond to the atomic number and atomic mass of the target
nucleus, respectively. For individual elastic $N$-$p$ and $N$-$n$
cross sections one obtains
\begin{eqnarray}
\sigma_{\rm SI}^{N-p,n}=\delta_Z^2 \,\frac{g_2^4}{16\pi
c_W^4}\,\frac{\mu^2}{m_Z^4}\, (c_V^{p,n})^2\,, \qquad
c_V^p=1-4c_W^2\,, \quad c_V^n=-1\,,
\end{eqnarray}
where $c_V^{p,n}$ are the vector form factors of the proton and
neutron. The elastic T-neutron-nucleus cross section reads
\begin{eqnarray}
\sigma_{\rm SI}^{N-A} = \delta_Z^2 \,\frac{g_2^4}{16\pi
c_W^4}\,\frac{\mu^2}{m_Z^4}\, \Big[ Z\, c_V^p + (A-Z)\, c_V^n
\Big]^2\,.
\end{eqnarray}
For practical use, it is instructive to represent the
T-neutron-nucleon cross sections in numerical form, i.e.
\begin{eqnarray}
\sigma_{\rm SI}^{N-p}=1.5\times 10^{-40}\, {\rm cm}^2\, \times
\delta_Z^2 \, \left(\frac{\mu}{m_p}\right)^2\,, \qquad \sigma_{\rm
SI}^{N-n}=2.5\times 10^{-38}\, {\rm cm}^2\, \times \delta_Z^2 \,
\left(\frac{\mu}{m_n}\right)^2\,.
\end{eqnarray}

These cross sections are rather large and strongly constrained by
direct detection experiments. At the moment, XENON100
\cite{Aprile:2012nq} provides the most stringent limit on
$\sigma_{\rm SI}^{\rm nucleon}$ (per nucleon in the case of a xenon
target) for high mass Dirac T-neutrons which is roughly
\begin{eqnarray}
-\log_{10}\Big(\frac{\sigma_{\rm SI}^{\rm nucleon}}{{\rm
cm}^2}\Big)\simeq 44.6 - 43.4\,,
\end{eqnarray}
corresponding to the T-neutron mass range of about
\begin{eqnarray}
M_{B_{\rm T}}\simeq 0.1 - 2\, {\rm TeV}\,,
\end{eqnarray}
respectively. This limit immediately provides a strong bound on
Dirac vector-like T-neutron coupling to the $Z$-boson, namely,
\begin{eqnarray}
\delta_Z\lesssim 2\times 10^{-3}\,, \qquad  M_{B_{\rm T}}\lesssim
2\, {\rm TeV}\,, \label{delta-Z}
\end{eqnarray}
and a little bit weaker constraint for $M_{B_{\rm T}}> 2\, {\rm
TeV}$. This means that if the corresponding direct DM detection
limits are confirmed the vector-like Dirac T-baryons with standard
EW interactions, i.e. model I with $\delta_Z=1$ introduced above in
Eq.~(\ref{scenarios}), are firmly ruled out. The only possibility to
accommodate the Dirac T-baryons is in the model II with $\delta_Z\ll
1$ which, however, requires an extra assumption about the existence
of an extra vector $SU(2)_{\rm V}\not=SU(2)_{\rm W}$ gauge symmetry
in the T-quark/T-baryon sector. The constraint (\ref{delta-Z}) is
then the upper limit on the mixing parameter (e.g. sine of a mixing
angle) between $Z$ and new $Z'$ bosons from an unspecified
high-scale $SU(2)_{\rm V}$ which should also be additionally
constrained by EW precision tests and extra gauge bosons ($Z'$ and
$W'$) searches at the LHC. The latter analysis can therefore be
performed together with the direct DM detection limits which,
however, goes beyond the present scope.

\section{Summary and conclusions}

In this work we have investigated basic properties of the Dirac
vector-like T-neutron DM predicted by the vector-like Technicolor
model \cite{our-TC} and have found important limitations on the
structure of T-strong dynamics from the direct DM detection data (in
particular, by the XENON100 experiment \cite{Aprile:2012nq}). This
has been done in the simplest QCD-like setting of the T-confinement
$SU(3)_{\rm TC}$ group and one generation of T-quarks. We have shown
that under a natural assumption about the T-baryon number
conservation, the local chiral symmetry breaking gives rise to the
vector $SU(2)_{\rm V}$ gauge symmetry, which acts on T-quark and
T-baryons sectors additional to the SM sectors. Whether the
$SU(2)_{\rm V}$ group is identified with the weak isospin
$SU(2)_{\rm W}$ symmetry of the SM or not provides us with the two
possible models I and II, respectively, for the structure of
(vector) weak interactions in the T-quark/T-baryon sectors which
have been studied above in detail.

As a crucial specific prediction of the vector-like TC model, the
T-quark and hence the T-baryon mass spectrum is degenerated at tree
level. The EW radiative corrections with vector T-baryon-$Z,\gamma$
couplings effectively split the T-baryon sector making the T-proton
$P=(UUD)$ (slightly) heavier than the T-neutron $N=(UDD)$
irrespectively of the nature of $SU(2)_{\rm V}$ gauge group. The
lightest T-baryon state is then the T-neutron which provides us with
the prominent heavy self-interacting DM candidate with many
appealing features. Two scenarios with symmetric and partially
asymmetric T-neutron DM have been considered and limits on T-baryon
mass scale have been derived from the DM relic abundance data.
Together with the naive QCD scaling hypothesis, this provides an
effective lower bound on the new T-confinement scale in a few TeV
range.

As was discussed thoroughly in Ref.~\cite{our-TC}, the EW precision
constraints at the fundamental level can only be satisfied for the
vector-like T-quarks under the weak isospin $SU(2)_{\rm W}$.
Alternatively, one could introduce the vector-like weak interactions
via a small mixing $\delta_Z\ll 1$ which can be limited to not upset
the SM tests. While both scenarios can be satisfied by all existing
EW and collider constraints, it turned out that only the latter
scenario with QCD-type TC group can be consistent with the DM
astrophysics constraints. This provides an extra very important
limit on the structure of TC sectors and new strongly-coupled
dynamics.

Indeed, the Dirac T-baryons were originated under the simplest
assignment of an odd T-confined group in the T-quark sector having
rank three, i.e. $SU(3)_{\rm TC}$. Thus, we conclude that in this
case or, more generally, in the case of any odd group $SU(2n+1)_{\rm
TC},\, n=1,2,\dots$, it is not possible to introduce the standard EW
interactions over the SM $SU(2)_{\rm W}$ gauge group in a
phenomenologically consistent way as suggested by the XENON100
constraint (\ref{delta-Z}). The only way to satisfy the existing
phenomenological (EW, collider and astrophysics) constraints is to
consider an even T-confinement $SU(2n)_{\rm TC},\, n=1,2,\dots$
group, for example, the simplest $SU(2)_{\rm TC}$ where the lightest
stable neutral T-baryon state $B_0=(UD)$ is scalar and does not
interact with the $Z$-boson thus evading the direct detection
limits. Corresponding analysis is ongoing.

\vspace{1cm} {\bf Acknowledgments}

Stimulating discussions and helpful correspondence with Johan
Bijnens, Sabir Ramazanov, Johan Rathsman, Francesco Sannino and
Torbj\"orn Sj\"ostrand are gratefully acknowledged. This work was
supported by the Crafoord Foundation (Grant No. 20120520). R. P. is
grateful to the ``Beyond the LHC'' Program at Nordita (Stockholm)
for support and hospitality during completion of this work.



\end{document}